\documentclass[11pt,a4paper]{article}

\usepackage[margin=1.2in]{geometry}
\usepackage{amsmath,amssymb,mathtools}
\usepackage{color}
\usepackage{graphics}
\usepackage{tikz}
\usepackage{float}
\usepackage[title,toc,titletoc,header]{appendix}
\usepackage{multirow}
\usepackage{url}
\usepackage{authblk}

\usetikzlibrary{angles,quotes}
\usetikzlibrary{calc}
\usetikzlibrary{arrows,decorations.markings}

\DeclareMathOperator*{\argmin}{arg\,min}

\newcommand{\SLS}{SLS}

\newcommand{\xddots}{%
	\raise 4pt \hbox {.}
	\mkern 6mu
	\raise 1pt \hbox {.}
	\mkern 6mu
	\raise -2pt \hbox {.}
}

\newtheorem{theorem}{Theorem}
\newtheorem{remark}{Remark}
\newtheorem{lemma}{Lemma}
\newtheorem{corollary}{Corollary}
\newtheorem{definition}{Definition}
\newtheorem{proposition}{Proposition}

\providecommand{\keywords}[1]
{
	\small	
	\textbf{\textit{Keywords---}} #1
}

\title{Learning stability of partially observed switched linear systems}

\author[1]{Zheming Wang \thanks{wangzheming@zjut.edu.cn}}
\author[2,3]{Rapha\"el M. Jungers \thanks{raphael.jungers@uclouvain.be. Rapha\"el M. Jungers is a  FNRS honorary Research Associate. This project has received funding from the European Research Council (ERC) under the European Union's Horizon 2020 research and innovation programme under grant agreement No 864017 - L2C. Rapha\"el M. Jungers is also supported by the Innoviris Foundation and the FNRS (Chist-Era Druid-net). He is currently on sabbatical leave at Oxford University,Department of Computer Science, Oxford, UK.}}
\author[4]{Mih\'{a}ly Petreczky \thanks{mihaly.petreczky@centralelille.fr}}
\author[1]{Bo Chen \thanks{bchen@zjut.edu.cn}}
\author[1]{Li Yu \thanks{lyu@zjut.edu.cn}}

\affil[1]{The Department of Automation, Zhejiang University of Technology, Hangzhou 310023, China}
\affil[2]{The ICTEAM Institute, UCLouvain, Louvain-la-Neuve,1348, Belgium}
\affil[3]{The Department of Computer Sciences, University of Oxford, OX1 3QD Oxford, United Kingdom}
\affil[4]{The Centre de Recherche en Informatique, Signal et Automatique de Lille, UMR CNRS 9189, CNRS; Ecole Centrale Lille, Université de Lille,  Villeneuve dAscq 59651, France}

\date{}

\begin{document}
	\maketitle

\begin{abstract}
	This paper deals with learning stability of partially observed switched linear systems under arbitrary switching. Such systems are widely used to describe cyber-physical systems which arise by combining physical systems with digital components. In many real-world applications, the internal states cannot be observed directly. It is thus more realistic to conduct system analysis using the outputs of the system. Stability is one of the most frequent requirement for safety and robustness of cyber-physical systems. Existing methods for analyzing
	stability of switched linear systems often require the knowledge of the parameters and/or all the states of the underlying system.
	In this paper, we propose an algorithm for deciding stability 
	of switched linear systems under arbitrary switching based purely on  observed output data. 	The proposed algorithm essentially relies on an output-based Lyapunov stability framework and returns an estimate of the joint spectral radius (JSR). We also prove a probably approximately correct error bound on the quality of the estimate of the JSR from the perspective of statistical learning theory.
\end{abstract}

\section{Introduction}
Verification of safety and robustness of AI systems has gained significant attention in recent years \cite{AIreport,dietterich2019robust,hamon2020robustness}. This is particularly important in the context of application of machine learning algorithms to \emph{cyber-physical systems} (CPS), which tend to be safety-critical (e.g., autonomous vehicles, etc.), see \cite{CyberPhysical,CyberPhysical3}. On the other hand, stability is one of the most basic requirements for safety and robustness of CPS
\cite{CyberPhysical,CyberPhysical3,HybSysNature}. Due to the interactions between the cyber and physical components, CPSs can be modelled by so called hybrid systems \cite{CyberPhysical3,HybSysNature,teel}, of which switched linear systems (SLS) represent a subclass \cite{BOO:L03,Sun:Book}. It is thus of interest to investigate verification of stability of SLSs. 

The topic of stability is a classical one in dynamical systems theory and in 
control theory  \cite{Son:MathContr,hirsch2012differential}, and has been studied for SLSs \cite{teel,BOO:L03,Sun:Book}. In the above-cited works, stability analysis techniques typically require a dynamical model of the system. However, obtaining an accurate model of real-world engineering systems by first principles can be challenging in particular for CPSs with hybrid behaviors. From the point of view of conventional control theory, system identification where dynamical models are learned from data could be a reasonable intermediate step for system analysis. In fact, system identification is a well-studied subject with a huge literature, see, e.g.,  \cite{INC:L98,ART:SL19} and the references therein. Early work mainly focuses on linear systems and tends to provide asymptotic guarantees see \cite{INC:L98} for an overview of the classical literature (including extensions to nonlinear systems). For hybrid systems, there also exists a rich literature on learning SLS models, see \cite{IdentSurvey,BOO:LB18,VidalAutomatica,Bako20,ART:MLG22} and the references therein. Many of existing techniques assume the knowledge of the switching signal \cite{CoxAutomatica,Verhaegen09}. When the switching signal is unavailable, auto-regressive models are often used to approximate SLSs, which can be computationally expensive for an accurate approximation \cite{VidalAutomatica}. In fact, it is proved in \cite{ART:L16} that learning SLS models is NP-hard without the knowledge of the switching signal. In this paper, we address stability of SLS state-space representations with unobserved switching. Our goal is to decide stability of a SLS directly based only on the observed data generated by it. More precisely, we consider discrete-time SLSs with no inputs under arbitrary switching, and we
are interested in deciding if the system is stable, i.e., if the joint spectral radius (JSR) 
\cite{BOO:J09} of the system matrices is smaller
than $1$. Our motivation is to bypass the identification phase, in order to avoid both computational burden and potential modelling errors.

We assume that we can observe random samples of the output of
the system. Moreover, we assume that each sampled output trajectory is  generated from a certain
initial state and switching signal. By output we mean a function of the hidden internal state. We then formulate an optimization problem on the observed outputs, such that the solution of this optimization problem gives an upper bound on the JSR of the underlying system with a certain precision and with a certain probability. That is, the optimization problem gives an upper bound on the JSR with high probability. From the perspective of statistical learning theory \cite{BOO:V99,BOO:SB14}, our result provides a \emph{probably approximately correct}  (PAC) bound, which brings important insights into the relation between the size of the sample and the precision of the solution.

The papers which are the closest to the present one are \cite{ART:KBJT19,INP:BJW21,INP:RWJ21,ART:WJ21b}
where the celebrated scenario approach is applied in order to infer stability of the system from a sampled set of observations. They show that, even though it is not obvious that the scenario approach can be used in this context, the geometric properties of SLSs allow to retrieve firm, probabilistic, stability guarantees under a set of mild assumptions.  We refer the reader to \cite{ART:CC05,ART:CC06,ART:CG08,ART:C10,ART:MGL14} for good introductions to the scenario approach. In contrast to \cite{ART:KBJT19,INP:BJW21,INP:RWJ21,ART:WJ21b} where fully observed SLSs are considered, this paper considers partially observed SLSs, i.e., we no longer assume that the whole internal state can be observed. The extension from the fully observed case to the partially observed case allows us to consider a wider range of practical systems where the internal state is not directly accessible. However, the challenge is that we need to construct a Lyapunov function from the output trajectory instead of the state trajectory. More specifically, for this extension, we need to develop an output-based Lyapunov stability analysis technique which serves as a basis for the proposed data-driven approach under observability conditions. Similar to \cite{ART:KBJT19}, we show that the data-based solution converges to the model-based solution as the sample size increases. In addition, we explicitly derive a convergence rate which provides important insights into the relation between the sample size and the precision of the solution. With the obtained convergence rate, we also show that the output-based Lyapunov function is PAC learnable in the sense of Valiant's
definition in \cite{ART:V84} using the proposed algorithm.

Our work is also closely related with the research on learning safety certificates, such as Lyapunov functions and contraction metrics, which guarantee stability, see 
\cite{chen2020learning,boffi2021learning,taylor2020learning,ravanbakhsh2019learning,chang2019neural,richards2018lyapunov,tsukamoto2021contraction,chang2021stabilizing,dawson2022safe,giesl2020approximation}. The latter references require the knowledge of the underlying system, except
\cite{taylor2020learning,boffi2021learning,chang2021stabilizing,giesl2020approximation}. Moreover,all these works require the knowledge of the full state and hybrid behaviors are not taken into consideration. Hence, these techniques are not suitable for the stability analysis of partially observed switched systems.


The contribution of this paper is threefold. First, we propose a stability learning approach for SLSs with only partial observation under observability assumptions. Second, we provide a PAC bound on the JSR with an explicit convergence rate, which allows to derive the \emph{sample complexity} of the proposed learning approach. Third, we conduct a comparison with hybrid system identification techniques \cite{IdentSurvey,BOO:LB18} using numerical examples. More precisely, by the numerical experiments in Section \ref{sec:num}, we show that hybrid system identification techniques \cite{IdentSurvey,BOO:LB18} not only are computationally expensive but also generate modelling errors which leads to false stability inference.

The rest of the paper is organized as follows. In Section \ref{sec:problem}, we present the formal problem formulation and some preliminary results on stability of SLSs. Section \ref{sec:leaning} is the main part of the paper, where we present the proposed output-based Lyapunov framework from the stability learning problem to probabilistic guarantees. In Section \ref{practical:issue}, we discuss some practical issues about the proposed analysis. In Section \ref{sec:num}, we provide numerical simulations and comparison with hybrid system identification techniques.

\textbf{Notation}.	We denote by $\mathbb{R}^+$ and $\mathbb{Z}^{+}$ the set of all non-negative real numbers and the set of all non-negative integers respectively. For a square matrix $Q$, $Q \succ (\succeq) 0$ means $Q$ is positive definite (semi-definite). For a symmetric $Q\succ 0$, let $\kappa(Q) \coloneqq \lambda_{\max}(Q)/\lambda_{\min}(Q)$. Consider the set $\mathcal{M}\coloneqq\{1,2,\cdots,M\}$ for some given integer $M\in \mathbb{Z}^+$, $\mathcal{M}^k$ denotes the $k$-Cartesian product of $\mathcal{M}$ for any $k\in \mathbb{Z}^k$. Let $\pmb{\sigma}=(\sigma_0,\sigma_1,\cdots, \sigma_{k-1})$ be an element of $\mathcal{M}^k$. For any $a,b\in \mathbb{Z}^+$ with $b \ge a$, we denote the segment $(\sigma_a,\cdots,\sigma_b)$ by $\pmb{\sigma}_{a:b}$. 
For consistence, let $\pmb{\sigma}_{a:b}=\emptyset$ when $a>b$.  For any $p\ge 1$, the $p$-norm of a vector/matrix $x$ is $\|x\|_p$ ($\|x\|$ is the $2$-norm by default), and let $\|x\|_F$ denote the Frobenius norm.

	\section{Problem statement and preliminaries}
	\label{sec:problem}

	\noindent
	\textbf{Switched linear systems} 
	Below we will define the notion of switched linear systems and recall some basic properties of such systems. These properties will allow us to relate stability of the system with the observed behavior. 
	
	A discrete-time switched linear system (SLS) is a dynamical system with output of the form
	\begin{equation}
		x(t+1) = A_{\sigma(t)}x(t), ~ ~ ~ y(t) = C_{\sigma(t)}x(t), \quad t\in \mathbb{Z}^+ 
		\label{eq:LSS}
	\end{equation}
	where $x(t)\in \mathbb{R}^n$ is the state vector,  $y(t)\in \mathbb{R}^p$ is the output and $\sigma: \mathbb{Z}^+\rightarrow \mathcal{M}\coloneqq\{1,2,\cdots,M\}$ is a time-dependent switching signal that indicates the current active mode of the system among $M$ possible modes in $\{A_1,A_2,\cdots, A_M\}$.
	We will use the tuple $\Sigma= \left(n,\{(A_i,C_i): i\in \mathcal{M}\} \right)$ to denote the switched linear system above. 
	
	Intuitively, a SLS is just a collection of linear dynamical systems defined on the same state-space. During the evolution of the SLS, one switches from one linear system to another according to the switching signal.
	For more details on switched systems see \cite{BOO:BNO03,Sun:Book}.

    Informaly, we would like to decide stability of \eqref{eq:LSS} based on a finite number of observed output data points. 
	In order to state the problem formally, first we will define below what we mean by stability. Then, we will explain our assumption on the data collection mechanism.
	
	\noindent
	\textbf{Stability} 
	Let us recall some basic stability results on discrete-time SLSs. We begin with the formal definition of asymptotic stability below.
	\begin{definition}[Asymptotic stability]
		The discrete-time SLS \eqref{eq:LSS} is asymptotically stable if for any initial state $x(0) \in \mathbb{R}^n$ and switching signal $\sigma: \mathbb{Z}^+\rightarrow \mathcal{M}$, $\lim_{t \rightarrow +\infty} x(t)=0$.
	\end{definition}
	To characterize this asymptotic stability property, we recall the concept of the joint spectral radius (JSR) \cite{BOO:J09} of a \SLS\ $\Sigma$:
	\begin{align}\label{eqn:rhoAK}
		\rho(\Sigma)\coloneqq \lim\limits_{k\rightarrow \infty} \max\limits_{\pmb{\sigma}\in \mathcal{M}^k}\|A_{\pmb{\sigma}} \|^{1/k}.
	\end{align}
	It is well known that the \SLS\ $\Sigma$ is asymptotically stable under arbitrary switching if and only if $\rho(\Sigma) < 1$, see, e.g., \cite{BOO:J09}. In general, computing the exact value of the JSR $\rho(\Sigma)$ is a difficult problem. Hence, in practice, we approximate $\rho(\Sigma)$  by computing lower and upper bounds within a Lyapunov framework,  where the template of the Lyapunov function  $V: \mathbb{R}^n  \times \mathcal{M}^{k} \rightarrow \mathbb{R}$ is specified for some $k\in \mathbb{Z}^+$. More precisely, given a Lyapunov function $V$ above, such that for any sequence $(i_1,\dots, i_k) \in \mathcal{M}^k$,
		we solve the following Lyapunov inequality for some choice of integers $T > 0$:
		\begin{align}
			& V(A_{i_{1}} \cdots A_{i_{T-k}}x,(i_{T-k+1} \cdots i_{T})) \le  \label{lyap:gamma} \\
			& \gamma^{2(T-k)} V(x,i_1,\cdots, i_k), \quad \forall (i_1,\ldots, i_{T})) \in \mathcal{M}^k,  \forall x\in \mathbb{R}^n \nonumber
		\end{align}
		where $\gamma \in \mathbb{R}^+$ and $V$ is parameterized by some variables. We refer the reader to \cite{BOO:J09} for some popular Lyapounov templates. In this paper, we consider quadratic Lyapunov functions in the form of $V(x,i_1,\cdots, i_k)=x^T\mathcal{P}_{(i_1,\cdots, i_k)}x$ for 
		some positive definite matrix $\mathcal{P}_{(i_1,\cdots, i_k)} \succ 0$.
	
    	\noindent
		\textbf{Data collection}
		We randomly generate multiple trajectories of  (\ref{eq:LSS}) where the initial state $x_0$ is uniform and i.i.d. in the unit sphere $\mathbb{S}_{n-1}$
		and the switching signal $\sigma_t$ is uniform and i.i.d. in $\mathcal{M}$ for any $t\in \mathbb{Z}^+$. Suppose we generate $N\in \mathbb{Z}^+$ trajectories of length $T\in \mathbb{Z}^+$, the sample is denoted as
		\begin{align}\label{eqn:omegaN}
			\omega_N\coloneqq \{(x_0^i,\pmb{\sigma}^i): i = 1,2,\cdots, N\}.
		\end{align}
		where $\pmb{\sigma}^i = (\sigma_0^i,\sigma_1^i,\cdots, \sigma_{T-1}^i)\in \mathcal{M}^{T}$. For each sampling pair, we measure the output trajectory data
		\begin{align}
			\begin{aligned}
				y^i_t=C_{\sigma_t^i} x^i_t, x^i_{t+1} =A_{\sigma_t^i}x_t^i, 0\le t\le T-1, \forall i
			\end{aligned}
		\end{align}
		where the subscript $t$ denotes the time instant and the superscript $i$ denotes each trajectory. The whole observed data set is denoted by
		\begin{align}
			\mathcal{D}_{obs} &\coloneqq \{ y^i_t: 0\le t\le T-1, i= 1,2,\cdots, N\}.
		\end{align}
		\textbf{Formal problem formulation}
		Suppose System \eqref{eq:LSS} is not known, but the observations $\mathcal{D}_{obs}$ are available and
		they correspond to the random sample  $\omega_N$ in \eqref{eqn:omegaN}.
		Find an estimate $\gamma^*(\omega_N)$ of the JSR $\rho(\Sigma)$. Note that we do not require the information on the switching signal.
		
		\section{Learning stability from output data}\label{sec:leaning}
		In this section, we present a data-driven stability analysis approach for \SLS s with partial observation.
		We start with the procedure to estimate the JSR of a SLS in the form of \eqref{eq:LSS}. We then present the
		probabilistic guarantees which describe the quality of that estimate.

		\subsection{Estimating JSR from data}
		\label{sec:leaning:estim}
		We propose to estimate the JSR and the corresponding Lyapunov function by solving an optimization
		problem based on observed data. In order to state the optimization problem,
		for each trajectory, given the observed data set $\mathcal{D}_{obs}$ and some $k\in \mathbb{Z}^+$, let us define the following time-series data
		\begin{align}\label{eqn:vikzki}
			v^i_k = \begin{pmatrix}
				y^i_0\\
				y^i_1\\
				\vdots\\
				y^i_{k-1}
			\end{pmatrix},
			z_k^i = \begin{pmatrix}
				y^i_{T-k}\\
				y^i_{T-k+1}\\
				\vdots\\
				y^i_{T-1}
			\end{pmatrix}, \forall i.
		\end{align}
		For notational convenience, let the time-series data set be denoted by, $\forall k \le T-1$,
		\begin{align}
			\mathcal{D}_k \coloneqq \{(v^i_k,z_k^i): i=1,2,\cdots, N\}.
		\end{align}
		Given the data set $\mathcal{D}_k$ for a sufficiently large $k\in \mathbb{Z}^+$, we 
		estimate the JSR by solving the following scenario (or sampled-based) program
		\begin{subequations}\label{eqn:scenariobreaking}
			\begin{align}
				&\min_{\gamma \ge 0, P} (\gamma,\|P\|_{F})\\
				\textrm{s.t. } &z^\top P z  \le \gamma^{2(T-k)}v^\top P v, \forall (v,z) \in \mathcal{D}_k,\label{eqn:zPzvgamma}\\
				&  I\preceq P \preceq \bar{\lambda}I.
			\end{align}
		\end{subequations}
	where the minimization is implemented in the \emph{lexicographic order} \footnote{The first component comes first: $(\gamma_1,\|P_1\|_F) < (\gamma_2,\|P_2\|_F) $ if $\gamma_1 < \gamma_2$ or else $\gamma_1=\gamma_2$ and $\|P_1\|_F < \|P_2\|_F$.} of the components of the objective function. Let the solution be denoted by $(\gamma^*_k(\omega_N),P^*_k(\omega_N))$ and and suppose that it is unique. 
		\emph{The quantity $\gamma^*_k(\omega_N)$ will be our estimate of the JSR of the unknown system
			\eqref{eq:LSS}}. If $\gamma^*_k(\omega_N)$ is sufficiently
		smaller than $1$,
	then we can conclude that the underlying system is stable, at least with high
probability. 
		

		%
		
		\noindent
		\textbf{Computational complexity}
		By fixing $\gamma$, the constraint \eqref{eqn:zPzvgamma} is linear, so that the optimization problem \eqref{eqn:scenariobreaking} can be solved efficiently using SDP solvers \cite{BOO:BV04} and bisection on $\gamma$. 
		The computational complexity of solving it is
		polynomial in the number of data points ( note that we do not need to know the number of modes). 
		In particular, solving \eqref{eqn:scenariobreaking} is expected to be
		much less computationally expensive than identifying the underlying SLS, which essentially has a complexity growing exponentially with the number of modes.
		This is supported by numerical results, see Table \ref{tab:time} in Section
		\ref{sec:num}.
		
		\noindent
		\textbf{Intuition}
		Intuitively, the solution of \eqref{eqn:scenariobreaking} results in a 
 Lypaunov function $V(x,(i_1,\cdots, i_k))$ which can be rewritten as a quadratic function of the outputs
		and which satisfies \eqref{lyap:gamma} with $\gamma \le \gamma^*_k(\omega_N)$.
		More precisely, given any $k\in \mathbb{Z}^+$, consider a switching sequence $\pmb{\sigma}=(\sigma_0,\sigma_1,\cdots, \sigma_{T-1}) \in \mathcal{M}^T$. 
		For any initial state 
		$x$ of \eqref{eq:LSS} and any integer  $k,\ell \in \mathbb{Z}^+$ with $k+\ell\le T$, define 
		\begin{equation}
			\label{eq:obs1}
			Y_{\pmb{\sigma},x,\ell,k}=\begin{pmatrix} y(\ell) \\ y(\ell+1) \\ \vdots \\ y(\ell+k-1) \end{pmatrix} \\
		\end{equation}
		where $(y(0),\cdots y(T-1))$  is the output generated by \eqref{eq:LSS}
		for the switching signal $\sigma(i)=\sigma_i$, $i=0,\ldots,T-1$ and initial state $x(0)=x$.
		%
		%
		We consider quadratic Lyapunov functions of the output trajectory 
		which can be expressed as
		\begin{align}\label{eqn:yP}
			V(x,\pmb{\sigma})= Y_{\pmb{\sigma},x,0,k}^{\top} P Y_{\pmb{\sigma},x,0,k} 
		\end{align}	
		where $P \in \mathbb{R}^{kp\times kp}$ is a positive definite matrix. 
		
		It then follows that if $V$ satisfies \eqref{lyap:gamma}, then the constant $\gamma$ is an upper bound on the JSR of \eqref{eq:LSS} when $k$ and $T$ satisfy certain conditions, as we will see in Theorem \ref{thm:jsr}.  In turn, $V$ satisfies \eqref{lyap:gamma}, if it satisfies the following inequality:
		for every $\pmb{\sigma} \in \mathcal{M}^T$
		\begin{equation}
			\label{eq:int1.0}
			\begin{split}
				&  Y_{\pmb{\sigma},x,T-k,k}^{\top} P Y_{\pmb{\sigma},x,T-k,k} \le  \gamma^{2(T-k)} Y_{\pmb{\sigma},x,0,k}^{\top} P Y_{\pmb{\sigma},x,0,k}
			\end{split}
		\end{equation}
	for every initial state $x$.
	Since any output sequence $\left(y(0),y(1),\cdots, y(T-1)\right)$ arises
	as $Y_{\pmb{\sigma},x,0,T}$ for some initial state $x$ and switching $\pmb{\sigma} \in \mathcal{M}^T$, it follows that \eqref{eq:int1.0} is equivalent to
	requiring that 
	\begin{equation}
		\label{eq:int1}
		\begin{split}
			& \begin{pmatrix}  y(T-k)\\y(T-k+1) \\ \vdots\\y(T-1) \end{pmatrix}^\top P  
			\begin{pmatrix}  y(T-k)\\y(T-k+1) \\ \vdots\\y(T-1) \end{pmatrix} \\ \le 
		& \gamma^{2(T-k)}  \begin{pmatrix}  y(0)\\y(1) \\ \vdots\\y(k-1) \end{pmatrix}^\top P \begin{pmatrix}  y(0)\\ 
			y(1) \\ \vdots\\y(k-1) \end{pmatrix}
	\end{split}
\end{equation}
holds for any output sequence $\left(y(0),y(1),\cdots, y(T-1)\right)$  and generated by \eqref{eq:LSS}.
Hence, in order to find a bound of the JSR, we have to solve the optimization problem for some sufficiently large $T\in \mathbb{Z}^+$
\begin{equation}
	\label{eqn:gammaoPo:alt} 
	\begin{split}
		& (\gamma^o_k,P^o_k) \coloneqq \argmin_{\gamma \ge 0, P \succeq I} (\gamma,\|P\|_{F})\\
		&		\textrm{s.t. }  P, \gamma \textrm{ satisfy \eqref{eq:int1} for all output trajectories } \\
		&   \left(y(0),y(1),\cdots, y(T-1)\right)
	\end{split}		
\end{equation}
The scenario program \eqref{eqn:scenariobreaking} can be in fact considered as a sampled version of the robust optimization problem \eqref{eqn:gammaoPo:alt}.

\subsection{Pathwise observability}
Since our goal is to learn stability from the output trajectory, first
we need to recall 
some technical concepts on observability of switched systems. By incorporating these concepts into Lyapunov stability analysis, we present new results that are needed for developing formal probabilistic stability guarantees in the sequel.

\begin{definition}\label{def:obssig}
	A  switching sequence $\pmb{\sigma} \in \mathcal{M}^{k}$  is said to
	be observable for \eqref{eq:LSS}, if the following implication
	holds: 
	$$Y_{\pmb{\sigma},x,0,k}=0 \implies x=0, $$
	where $Y_{\pmb{\sigma},x,0,k}$ is as in \eqref{eq:obs1}.
	The smallest $k\in \mathbb{Z}^+$ such that there exists an 
	observable switching sequence
	$\pmb{\sigma}\in \mathcal{M}^{k}$ is called the \emph{observability index} of $\Sigma$, denoted by $h(\Sigma)$.
\end{definition}
Intuitively, a switching signal is observable, then the state of the system can be reconstructed from the observed outputs.
When a \SLS\  $\Sigma$ has no observable switching signal, we let $h(\Sigma)=\infty$. We also recall a stronger observability condition from \cite{babaali2003pathwise},  called \emph{pathwise observable}.
\begin{definition}\label{def:strongobs}
	The SLS \eqref{eq:LSS} is said pathwise observable if there exists $k\in \mathbb{Z}^+$ such that  
	every switching signal $\pmb{\sigma} \in \mathcal{M}^k$ is observable. We refer to the smallest such integer $k$ as the \emph{pathwise observability index}, denoted by $\mathcal{H}(\Sigma)$.
\end{definition}
Intuitively, \emph{pathwise observability} means that the state of the system can be  reconstructed from the observed outputs, for any choice of the switching signal.
When an \SLS\ $\Sigma$ is not \emph{pathwise observable}, we say that $\mathcal{H}(\Sigma) = \infty$. Let us also point out that, the condition of \emph{pathwise observable} is decidable as shown in \cite{babaali2003pathwise,ART:JKH17} by providing explicit upper bounds on $\mathcal{H}(\Sigma)$.

For any $a,b\in \mathbb{Z}^+$, let us also define  $A_{\pmb{\sigma}_{a:b}}\coloneqq A_{\sigma_b} \cdots A_{\sigma_a}$. By convention, when $a>b$, let $A_{\pmb{\sigma}_{a:b}} = I$. Following \cite{INP:BE04}, for any $\pmb{\sigma}=(\sigma_0,\sigma_1,\cdots, \sigma_{k-1}) \in \mathcal{M}^k$ of length $k\in \mathbb{Z}^+$, we define the path-dependent observability matrix
\begin{align}
	\mathcal{O}_{\Sigma}(\pmb{\sigma}) \coloneqq \begin{pmatrix}
		C_{\sigma_0}\\
		C_{\sigma_1}A_{\sigma_0}\\
		\vdots\\
		C_{\sigma_{k-1}}A_{\sigma_{k-2}} \cdots A_{\sigma_1}A_{\sigma_0}
	\end{pmatrix}.
\end{align}

\begin{remark}
	Using the definition of path-dependent observability matrices, a switching sequence $\pmb{\sigma} \in \mathcal{M}^{k}$ for some $k\in \mathbb{Z}^+$ is said to be observable if $\textrm{rank} \left(\mathcal{O}_{\Sigma}(\pmb{\sigma})\right) = n$. Following this, we say that a \SLS\ $\Sigma= \left(n,\{(A_i,C_i): i\in \mathcal{M}\} \right)$ is pathwise observable if there exists $k\in \mathbb{Z}^+$ such that $\textrm{rank}\left(\mathcal{O}_{\Sigma}(\pmb{\sigma})\right) = n$ for any $\pmb{\sigma} \in \mathcal{M}^k$.
\end{remark}

Note that the output trajectory can be rewritten as
$
Y_{\pmb{\sigma},x,0,k} = 	\mathcal{O}_{\Sigma}(\pmb{\sigma})x
$, the quadratic Lyapunov function in \eqref{eqn:yP} can be rewritten as
$
V(x,\pmb{\sigma}) = x^\top \mathcal{O}_{\Sigma}(\pmb{\sigma})^\top  P\mathcal{O}_{\Sigma}(\pmb{\sigma})x.
$
For notational convenience, we then define the following matrices based on the switching sequence 
\begin{align}\label{eqn:Pellk}
	\mathcal{P}^{\pmb{\sigma}}_\Sigma(P,\ell,k) = \mathcal{O}_{\Sigma}(\pmb{\sigma}_{\ell:k+\ell-1}) ^\top P\mathcal{O}_{\Sigma}(\pmb{\sigma}_{\ell:k+\ell-1}).
\end{align}

With the definitions above, we obtain a Lyapunov stability result for  SLSs.

\begin{theorem}\label{thm:jsr}
	Consider an SLS, denoted by $\Sigma= \left(n,\{(A_i,C_i): i\in \mathcal{M}\} \right)$, suppose that $\Sigma$ admits at least one observable switching signal, and denote $h(\Sigma)$ the observability index as in Definition \ref{def:obssig}. Assume that there exist $k \in \mathbb{Z}^+$, $\ell \in \mathbb{Z}^+$, $\gamma \ge 0$ and $P \succ 0$ such that $k\ge h(\Sigma)$, and for any $\pmb{\sigma} \in \mathcal{M}^{k+\ell}$, 
	\begin{align}\label{eqn:AOPgamma}
		A_{\pmb{\sigma}_{0:\ell-1}}^\top \mathcal{P}^{\pmb{\sigma}}_\Sigma(P,\ell,k)A_{\pmb{\sigma}_{0:\ell-1}} \preceq \gamma^{2\ell} \mathcal{P}^{\pmb{\sigma}}_\Sigma(P,0,k),
	\end{align}
	where $ \mathcal{P}^{\pmb{\sigma}}_\Sigma(P,0,k)$ and $\mathcal{P}^{\pmb{\sigma}}_\Sigma(P,\ell,k)$ are given as in \eqref{eqn:Pellk}. Then, $\rho(\Sigma) \le \gamma $.
\end{theorem}
Proof: For any $q\in \mathbb{Z}^+$ and $\pmb{\sigma} \in \mathcal{M}^{q\ell+k}$, from  (\ref{eqn:AOPgamma}), it holds that
\begin{align*}
	&A_{\pmb{\sigma}_{(q-1)\ell:q\ell-1}}^\top \mathcal{P}^{\pmb{\sigma}}_\Sigma(P,q\ell,k) A_{\pmb{\sigma}_{(q-1)\ell:q\ell-1}} \\
	\preceq &\gamma^{2\ell} \mathcal{P}^{\pmb{\sigma}}_\Sigma(P,(q-1)\ell,k). 
\end{align*}
This implies that
\begin{align}
	&A_{\pmb{\sigma}_{0:q\ell-1}}^\top \mathcal{P}^{\pmb{\sigma}}_\Sigma(P,q\ell,k)  A_{\pmb{\sigma}_{0:q\ell-1}} \nonumber\\
	\preceq 
	& \gamma^{2\ell} A_{\pmb{\sigma}_{0:(q-1)\ell-1}}^\top\mathcal{P}^{\pmb{\sigma}}_\Sigma(P,(q-1)\ell,k)  A_{\pmb{\sigma}_{0:(q-1)\ell-1}} \nonumber\\
	& \vdots \nonumber\\
	\preceq 
	& \gamma^{2q\ell} \mathcal{P}^{\pmb{\sigma}}_\Sigma(P,0,k),\quad   \forall \pmb{\sigma} \in \mathcal{M}^{q\ell+k}, q\in \mathbb{Z}^+. \label{eqn:APqlok}
\end{align}
Since $k\ge h(\Sigma)$, there exists $\tilde{\pmb{\sigma}} \in \mathcal{M}^k$ such that $rank\left( \mathcal{O}_{\Sigma}(\tilde{\pmb{\sigma}}) \right) = n$ from Definition \ref{def:obssig}. Thus, there exist $\overline{c} \ge \underline{c} > 0$ such that $ \mathcal{O}_{\Sigma}(\tilde{\pmb{\sigma}})^\top \mathcal{O}_{\Sigma}(\tilde{\pmb{\sigma}}) \succeq \underline{c} I$ and $ \mathcal{O}_{\Sigma}(\pmb{\sigma})^\top \mathcal{O}_{\Sigma}(\pmb{\sigma})\preceq \overline{c} I$ for any $\pmb{\sigma} \in \mathcal{M}^k$. With this, we have that, for any $\pmb{\sigma} \in \mathcal{M}^{q\ell+k}$ with $\pmb{\sigma}_{q\ell:k+q\ell-1} = \tilde{\pmb{\sigma}}$, 
\begin{align*}
	A_{\pmb{\sigma}_{0:q\ell-1}}^\top \mathcal{P}^{\pmb{\sigma}}_\Sigma(P,q\ell,k)  A_{\pmb{\sigma}_{0:q\ell-1}} &\succeq \lambda_{\min}(P) \underline{c} A_{\pmb{\sigma}_{0:q\ell-1}}^\top A_{\pmb{\sigma}_{0:q\ell-1}},\\
	\mathcal{P}^{\pmb{\sigma}}_\Sigma(P,0,k) &\preceq \overline{c} \lambda_{\max}(P) I.
\end{align*}
The two inequalities above, together with \eqref{eqn:APqlok}, imply that  
\begin{align*}
\lambda_{\min}(P) \underline{c} A_{\pmb{\sigma}_{0:q\ell-1}}^\top A_{\pmb{\sigma}_{0:q\ell-1}} \preceq \gamma^{2q\ell} \overline{c} \lambda_{\max}(P) I
\end{align*} 
which means that
\begin{align}
	\|A_{\pmb{\sigma}}\| \le \gamma^{q\ell} \sqrt{ \kappa(P)\frac{\overline{c}}{\underline{c}}}, \quad \forall \pmb{\sigma} \in \mathcal{M}^{q\ell}
\end{align}
Hence, 
\begin{align*}
	\rho(\Sigma) = \lim\limits_{q\rightarrow \infty} \max\limits_{\pmb{\sigma}\in \mathcal{M}^{ql}}\|A_{\pmb{\sigma}} \|^{\frac{1}{ql}}\le \gamma \lim\limits_{q\rightarrow \infty} \left( \kappa(P)\frac{\overline{c}}{\underline{c}}\right)^{\frac{1}{2ql}} =\gamma.
\end{align*}
$\Box$

The stability result in Theorem \ref{thm:jsr} serves as a basis for the rest of the paper. Observe that the constraint \eqref{eqn:AOPgamma} is not amenable as it is for a data-driven approach, as it explicitly uses the system matrices. In order to leverage it in a data-driven framework, we now provide an interpretation of this stability condition from the perspective of robust optimization. Suppose the overall length of the output trajectory is $T$. For any $k \in \mathbb{Z}^+$, we formulate the following robust optimization problem
\begin{subequations}\label{eqn:gammaoPo}
	\begin{align}
		(\gamma^o_k,&P^o_k) \coloneqq \argmin_{\gamma \ge 0, P \succeq I} (\gamma,\|P\|_{F})\\
		\textrm{s.t. } & x^\top A_{\pmb{\sigma}_{0:\ell-1}}^\top \mathcal{P}^{\pmb{\sigma}}_\Sigma(P,T-k,k)A_{\pmb{\sigma}_{0:\ell-1}}x \label{eqn:xAPAx}\\
		&\le \gamma^{2(T-k)} x^\top \mathcal{P}^{\pmb{\sigma}}_\Sigma(P,0,k)x, \forall (x,\pmb{\sigma})\in \mathbb{S}_{n-1}\times \mathcal{M}^{T}  \nonumber
	\end{align}
\end{subequations}	
where $ \mathcal{P}^{\pmb{\sigma}}_\Sigma(P,0,k)$ and $\mathcal{P}^{\pmb{\sigma}}_\Sigma(P,\ell,k)$ are given as in \eqref{eqn:Pellk}. By homogeneity, \eqref{eqn:xAPAx} is equivalent to \eqref{eqn:AOPgamma}. From this robust optimization formulation, the stability condition in Theorem \ref{thm:jsr} means that for any initial state $x(0)$ the aggregated quadratic Lyapunov function in \eqref{eqn:yP} is decreasing at a rate of $\gamma$. The problem \eqref{eqn:gammaoPo} can be also seen as a reformulation of \eqref{eqn:gammaoPo:alt}

If all the dynamics matrices are nonsingular, Theorem \ref{thm:jsr} in fact implies the \emph{pathwise observability} property, as stated in the following corollary.
\begin{corollary}\label{cor:strongobs}
	Suppose the conditions in Theorem \ref{thm:jsr} hold. Suppose the matrices $\{A_i\}_{i\in \mathcal{M}}$, $\ell \ge k$, and $\gamma >0$, the stability condition (\ref{eqn:AOPgamma}) also implies that $\textrm{rank}\left(\mathcal{O}_{\Sigma}(\pmb{\sigma})\right) = n$ for any $\pmb{\sigma} \in \mathcal{M}^k$.
\end{corollary}
Proof: Since $\ell \ge k$, there is no overlapping between  $\pmb{\sigma}_{0:k-1}$ and  $\pmb{\sigma}_{l:k+l-1}$. Again, let $\tilde{\pmb{\sigma}} \in \mathcal{M}^k$ be such that $rank\left( \mathcal{O}_{\Sigma}(\tilde{\pmb{\sigma}}) \right) = n$. With the fact that $A_i$ is invertible for any $i\in \mathcal{M}$ (as $\Sigma$ is reversible), the condition (\ref{eqn:AOPgamma}) implies that, for any $\pmb{\sigma} \in \mathcal{M}^{k+\ell}$ with $\pmb{\sigma}_{l:k+l-1} = \tilde{\pmb{\sigma}}$,
\begin{align}
	\mathcal{P}^{\pmb{\sigma}}_\Sigma(P,0,k) \succeq  \gamma^{-2\ell}	A_{\pmb{\sigma}_{0:\ell-1}}^\top \mathcal{P}^{\pmb{\sigma}}_\Sigma(P,\ell,k)A_{\pmb{\sigma}_{0:\ell-1}} \succ 0.
\end{align}
This means that $\mathcal{O}_{\Sigma}(\pmb{\sigma})$ is full-column rank for any $\pmb{\sigma} \in \mathcal{M}^k$.
$\Box$

\begin{remark}
	From Corollary \ref{cor:strongobs}, we can see that \emph{pathwise observability} is not a conservative condition in our output-based Lyapunov framework in which the switching signal is not available.
\end{remark}

\subsection{Almost Lyapunov stability}
Before we prove our main result, we  introduce the concept of almost Lyapunov stability which allows to describe the stability with unseen regions in the state space. More precisely, we aim to derive formal guarantees on the JSR from an almost stability condition in which the aggregated Lyapunov functions based on the matrices as defined in \eqref{eqn:Pellk} are non-increasing everywhere except on a small subset whose measure is bounded by $\varepsilon\in (0,1)$.

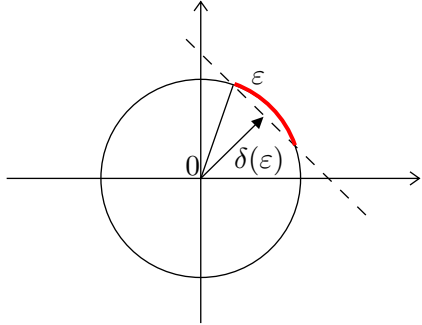
\begin{figure}[h]
	\centering
	\tikzset{every picture/.style={line width=0.5pt}} 
	\begin{tikzpicture}[x=0.5pt,y=0.5pt,yscale=-1,xscale=1]
		
		\draw  (173.67,191) -- (482.67,191)(319,57.67) -- (319,300.17) (475.67,186) -- (482.67,191) -- (475.67,196) (314,64.67) -- (319,57.67) -- (324,64.67)  ;
		\draw   (244.25,191) .. controls (244.25,149.72) and (277.72,116.25) .. (319,116.25) .. controls (360.28,116.25) and (393.75,149.72) .. (393.75,191) .. controls (393.75,232.28) and (360.28,265.75) .. (319,265.75) .. controls (277.72,265.75) and (244.25,232.28) .. (244.25,191) -- cycle ;
		\draw [color={rgb, 255:red, 0; green, 0; blue, 0 }  ,draw opacity=1 ] [dash pattern={on 4.5pt off 4.5pt}]  (308,86) -- (378.52,155.49) -- (444,220) ;
		\draw [color={rgb, 255:red, 0; green, 0; blue, 0 }  ,draw opacity=1 ]   (363.87,146.61) -- (319,191) ;
		\draw [shift={(366,144.5)}, rotate = 135.31] [fill={rgb, 255:red, 0; green, 0; blue, 0 }  ,fill opacity=1 ][line width=0.08]  [draw opacity=0] (8.93,-4.29) -- (0,0) -- (8.93,4.29) -- cycle    ;
		\draw    (343.26,120.74) -- (319,191) ;
		\draw  [draw opacity=0][line width=1.5]  (344.29,119.7) .. controls (365.44,127.4) and (382.17,144.49) .. (389.58,166) -- (319,191) -- cycle ; \draw  [color={rgb, 255:red, 255; green, 0; blue, 0 }  ,draw opacity=1 ][line width=1.5]  (344.29,119.7) .. controls (365.44,127.4) and (382.17,144.49) .. (389.58,166) ;  
		
		\draw (305.5,172.4) node [anchor=north west][inner sep=0.75pt]    {$0$};
		\draw (355,108.4) node [anchor=north west][inner sep=0.75pt]    {$\varepsilon $};
		\draw (342,164.4) node [anchor=north west][inner sep=0.75pt]    {$\delta ( \varepsilon )$};
		
	\end{tikzpicture}
	\caption{Illustration of $\delta(\varepsilon)$: $\varepsilon$ is the uniform (probability) measure of the spherical cap in red and $\delta(\varepsilon)$ is the distance to the base of the spherical cap.}\label{fig:delta}
\end{figure}

To formally state the stability result with an almost Lyapunov function, the following definition is also needed. Given any $\varepsilon\in (0,1)$, let
\begin{align}\label{eqn:delta}
		\delta(\varepsilon) \coloneqq
	\begin{cases} 
			\sqrt{1-\mathcal{I}^{-1}(2\varepsilon;\frac{n-1}{2},\frac{1}{2}}) &
			\varepsilon \in [0,\frac{1}{2}) \\
			0 & \varepsilon \ge \frac{1}{2}
		\end{cases}
\end{align}
where $\mathcal{I}(x;a,b)$ the regularized incomplete beta function defined as
\begin{align}\label{eqn:Ixab}
	\mathcal{I}(x;a,b) \coloneqq \frac{\int_0^x t^{a-1}(1-t)^{b-1}dt}{\int_0^1 t^{a-1}(1-t)^{b-1}dt}. 
\end{align}
A geometric interpretation of the function $\delta(\cdot)$  is given in Figure \ref{fig:delta}.

With this definition, we derive the following stability result under the Lyapunov condition \eqref{eqn:AOPgamma} 
with a violating subset.
\begin{theorem}\label{thm:jsrS}
	Consider a pathwise observable \SLS\ $\Sigma$ as in \eqref{eq:LSS}.
	Suppose there exist $k \in \mathbb{Z}^+$, $\ell \in \mathbb{Z}^+$, $\gamma \ge 0$, $P \succ 0$ and a subset $S\subseteq \mathbb{S}_{n-1}$ such that, $k\ge \mathcal{H}(\Sigma)$, and for any $\pmb{\sigma} \in \mathcal{M}^{k+\ell}$, 
	\begin{align}\label{eqn:AOPgammax}
		x^\top A_{\pmb{\sigma}_{0:\ell-1}}^\top \mathcal{P}^{\pmb{\sigma}}_\Sigma(P,\ell,k)A_{\pmb{\sigma}_{0:\ell-1}} x 
		\le  &\gamma^{2\ell}x^\top  \mathcal{P}^{\pmb{\sigma}}_\Sigma(P,0,k) x, \nonumber\\
		&\quad \forall x\in \mathbb{S}_{n-1}\setminus S.
	\end{align}
	where $ \mathcal{P}^{\pmb{\sigma}}_\Sigma(P,0,k)$ and $\mathcal{P}^{\pmb{\sigma}}_\Sigma(P,\ell,k)$ are given as in \eqref{eqn:Pellk}. Then, 
	\begin{align}\label{eqn:jsrS}
		\rho(\Sigma) \le \frac{\gamma}{\sqrt[\ell]{\delta\left(\frac{\mu(S)\chi_{\Sigma}(P,k)}{2}\right)}}
	\end{align}
	where $\mu(\cdot)$ is the uniform probability measure  on the unit sphere $\mathbb{S}_{n-1}$, $\delta(\cdot)$ is defined in \eqref{eqn:delta} and
	\begin{align}
		\chi_{\Sigma}(P,k) &\coloneqq \max_{\pmb{\sigma}\in \mathcal{M}^k} 	\sqrt{\frac{\det( \mathcal{P}^{\pmb{\sigma}}_\Sigma(P,0,k))}{\lambda_{\min}( \mathcal{P}^{\pmb{\sigma}}_\Sigma(P,0,k))^{n}}}.
		\label{eqn:kappaSigma}
	\end{align}
\end{theorem}

The proof of Theorem \ref{thm:jsrS} is given in the appendix. From the definition of $\chi_{\Sigma}(P,k)$ in \eqref{eqn:kappaSigma},  \emph{pathwise observability} is crucial for the stability result in Theorem \ref{thm:jsrS} in the sense that it guarantees the boundedness of $\chi_{\Sigma}(P,k)$. From Theorem \ref{thm:jsrS},  more explicit stability guarantees can be also derived, as shown in the following corollary. 
\begin{corollary}\label{cor:muSapriori}
	Given the same conditions as in Theorem \ref{thm:jsrS}, it holds that 
	$\chi_{\Sigma}(P,k) \le \sqrt{(c_k\kappa(P))^{n-1}}$ and
	\begin{align}\label{eqn:jsrSc}
		\rho(\Sigma) \le \frac{\gamma}{\sqrt[\ell]{\delta\left(\frac{\mu(S)\sqrt{(c_k\kappa(P))^{n-1}}}{2}\right)}},
	\end{align}
	where 
	\begin{align*}
		\kappa(P) = \frac{ \lambda_{\max}(P)}{ \lambda_{\min}(P)} \textrm{ and }  c_k\coloneqq \max_{\pmb{\sigma}\in \mathcal{M}^k} \kappa(\mathcal{O}_{\Sigma}(\pmb{\sigma})^\top \mathcal{O}_{\Sigma}(\pmb{\sigma})) 
	\end{align*}
	Moreover, when $I \preceq P\preceq \bar{\lambda} I$ for some $\bar{\lambda}\ge 1$, 
	\begin{align}\label{eqn:jsrScla}
		\rho(\Sigma) \le \frac{\gamma}{\sqrt[\ell]{\delta\left(\frac{\mu(S)\sqrt{(c_k\bar{\lambda})^{n-1}}}{2}\right)}}.
	\end{align}
\end{corollary}
Proof: By definition, since $\Sigma$ is pathwise observable with $\mathcal{H}(\Sigma) \le k$, there exists $r > 0$ such that $  r I \preceq \mathcal{O}_{\Sigma}(\pmb{\sigma})^\top \mathcal{O}_{\Sigma}(\pmb{\sigma})\preceq r c_k I$ for any $\pmb{\sigma} \in \mathcal{M}^k$, which leads to the following inequalities:
\begin{align*}
	\lambda_{\max} (\mathcal{P}^{\pmb{\sigma}}_\Sigma(P,0,k)) &\le \lambda_{\max}(\mathcal{O}_{\Sigma}(\pmb{\sigma})^\top \mathcal{O}_{\Sigma}(\pmb{\sigma}))\lambda_{\max} (P)\\
	&\le r c_k \lambda_{\max} (P),\\
	\lambda_{\min} (\mathcal{P}^{\pmb{\sigma}}_\Sigma(P,0,k)) &\ge  \lambda_{\min}(\mathcal{O}_{\Sigma}(\pmb{\sigma})^\top \mathcal{O}_{\Sigma}(\pmb{\sigma}))\lambda_{\min} (P)\\
	& \ge r\lambda_{\min} (P). 
\end{align*}
Using these inequalities, we arrive at
\begin{align*}
	\sqrt{\frac{\det( \mathcal{P}^{\pmb{\sigma}}_\Sigma(P,0,k))}{\lambda_{\min}( \mathcal{P}^{\pmb{\sigma}}_\Sigma(P,0,k))^{n}}} &\le \sqrt{\frac{\lambda_{\max}( \mathcal{P}^{\pmb{\sigma}}_\Sigma(P,0,k))^{n-1}}{\lambda_{\min}( \mathcal{P}^{\pmb{\sigma}}_\Sigma(P,0,k))^{n-1}}} \\
	&\le  \sqrt{\left( c_k\kappa(P)\right)^{n-1}}, \forall \pmb{\sigma} \in \mathcal{M}^k.
\end{align*}
Hence, it holds that $\chi_{\Sigma}(P,k) \le \sqrt{\left( c_k\kappa(P)\right)^{n-1}}$. Putting this inequality into \eqref{eqn:jsrS},  we obtain (\ref{eqn:jsrSc}). Note that the function $\delta(\cdot)$ is decreasing. When $\kappa(P)\le \bar{\lambda}$, (\ref{eqn:jsrSc}) implies (\ref{eqn:jsrScla}). $\Box$

With the results above, we are able to extend the technique in \cite{ART:KBJT19} which only considers the fully observed case to the partially observed case.

\subsection{Main result: Probabilistic stability certificates}\label{sec:probstability}
In the rest of this section, we formally present our main result, that is, stability certificates based on the solution of the scenario program 
\eqref{eqn:scenariobreaking}
in a probabilistic sense. 

Our derivation relies on the scenario approach (also known as scenario optimization) \cite{ART:CC05,ART:CC06,ART:C10}. In order to use \cite{ART:C10}, we need to fulfill a non-degeneracy assumption, formalized below.
\begin{definition}\label{def:degenerate}
	For the SLS \eqref{eq:LSS} 
	and $k,T\in \mathbb{Z}^+$, a switching sequence $\pmb{\sigma}\in \mathcal{M}^{T}$ is \emph{degenerate} if there exist $P\succ 0$ and $\gamma \ge 0$ such that 
	for any initial state $x$,
	\[ Y_{\pmb{\sigma},x,T-k,k}^{\top} P Y_{\pmb{\sigma},x,T-k,k} = \gamma^{2(T-k)} Y_{\pmb{\sigma},x,0,k}^{\top}  P Y_{\pmb{\sigma},x,0,k}  \]
	where $Y_{\pmb{\sigma},x,0,k}$,  $Y_{\pmb{\sigma},x,T-k,k}$
	are as in \eqref{eq:obs1}.
\end{definition}
With the discussions and definitions above, we are now ready to present the main result of this section.
\begin{theorem}[Main result]\label{thm:prostability}
	Consider the \SLS\ $\Sigma$ as given in \eqref{eq:LSS}, where the initial state is i.i.d with the uniform distribution over the unit sphere $\mathbb{S}_{n-1}$ and the switching signal is uniform and i.i.d. in $\mathcal{M}$. Given $N,T\in \mathbb{Z}^+$, the sample set $\omega_N$ as defined in \eqref{eqn:omegaN}, and $k\le T-1$, let $(\gamma_k^*(\omega_N),P_k^*(\omega_N))$ be the unique solution of the scenario program \eqref{eqn:scenariobreaking}. Assume that $\Sigma$ is pathwise observable with $\mathcal{H}(\Sigma) \le k$ and that none of the switching sequences in $ \mathcal{M}^{T}$ is degenerate in the sense of Definition \ref{def:degenerate}. For any $\epsilon \in (0,1)$, with probability no smaller than $1-\phi(\epsilon;d,N)$,
	\begin{align}\label{eqn:rhoSigQ}
		\rho(\Sigma) \le \frac{\gamma^*(\omega_N)}{\sqrt[T-k]{\delta\left(\frac{\epsilon M^T\chi_{\Sigma}(P^*(\omega_N),k)}{2}\right)}}
	\end{align}
	where $d=\frac{kp(kp+1)}{2}$,
	and $\chi_{\Sigma}(\cdot,k)$ is given in \eqref{eqn:kappaSigma}, $\delta(\cdot)$ is defined as in \eqref{eqn:delta}, and
	\begin{align} 
		\phi(\epsilon;d,N) &\coloneqq  1- \mathcal{I}(\epsilon;d,N-d+1). 
		\label{eqn:phiepsilon} 
	\end{align}
\end{theorem}

The proof of Theorem \ref{cor:complexity} is presented in the appendix. Let us emphasize that the assumption that none of the sequences is degenerate in Theorem \ref{thm:prostability} is not conservative in practice. In fact, this assumption is a particular case of Assumption 2 in \cite{ART:C10}, see Section 3.4 in \cite{ART:C10} for discussions on relaxing such an assumption. In the fully observed case, this assumption actually means that none of the matrices $\{A_i\}_{i=1}^M$ is similar to a Barabanov matrix, which is diagonalizable with all the eigenvalues having the same modulus, see \cite{INP:BJW21}. Let us also point out that the function $\chi_{\Sigma}(\cdot,k)$ as defined in \eqref{eqn:kappaSigma} implicitly depends on the parameters of the underlying system $\Sigma$. In Subsection \ref{expl:bounds} we present an algorithm for estimating $\chi_{\Sigma}(\cdot,k)$. 

The error bound of Theorem \ref{thm:prostability} implies that 
the true JSR is smaller than the sample-based solution $\gamma^*(\omega_N)$
multiplied with the correction factor 
$$\frac{1}{\sqrt[T-k]{\delta\left(\frac{\epsilon M^T\chi_{\Sigma}(P^*(\omega_N),k)}{2}\right)}}$$
with a high probability $1-\phi(\epsilon;d,N)$.
The probability $1-\phi(\epsilon;d,N)$ converges to $1$ as $N \rightarrow \infty$.
That is, the more data points we have, the more certain we are that \eqref{eqn:rhoSigQ} holds. 
The factor depends on the accuracy level $\epsilon$, and it tends to $1$ as $\epsilon \rightarrow 0$. We will explicitly discuss the convergence rate in Section \ref{sec:samplex}. 

The correction factor above is \emph{a posteriori}, in the sense that its numerical value depends on the outcome of the scenario program \eqref{eqn:scenariobreaking}. For this reason it may be difficult to evaluate the quality of this factor. Below we present a modified error bound where the factor with which
$\gamma^*(\omega_N)$  is multiplied does not depend on data. This then allows to bound
that factor using prior knowledge on the set of possible models. This is done by
exploiting the constraint $I\preceq P \preceq \bar{\lambda}I$ in \eqref{eqn:scenariobreaking}.  
\begin{corollary}\label{cor:complexity}
	Suppose that the conditions in Theorem \ref{thm:prostability} hold. There exists a constant $c\ge 1$, for any $\beta \in (0,1)$, with probability no smaller than $1-\beta$,
	\begin{align}\label{eqn:jsrapriori}
		\rho(\Sigma) \le \frac{\gamma_k^*(\omega_N) }{\sqrt[T-k]{\delta\left(\frac{\epsilon M^T\sqrt{(c\bar{\lambda})^{n-1}}}{2}\right)}} 
	\end{align}
	with $\epsilon = \mathcal{I}^{-1}(1-\beta;d,N-d+1)$ and $d=\frac{kp(kp+1)}{2}$.
\end{corollary}
Proof: This is a direct consequence of Theorem \ref{thm:prostability} and Corollary \ref{cor:muSapriori} with $c=c_k$. $\Box$

The constant $c$ in \eqref{eqn:jsrapriori} depends on the underlying system as shown in Corollary \ref{cor:muSapriori}.
Corollary \ref{cor:complexity} enables us to know \emph{a priori} how much data is needed to reach certain accuracy and confidence level.  That is, 
given a fixed confidence level $\beta$ and a fixed $\epsilon$, we are able to compute $N$ such that \eqref{eqn:jsrapriori} holds with probability $1-\beta$.
Note that this requires the knowledge of the
constant $c$, which depends on the matrices of the underlying system. 
Suppose that the underlying unknown system comes from a certain family of SLSs,
we can choose $c$ to be an upper bound on the constant $\max_{\pmb{\sigma}\in \mathcal{M}^k} \kappa(\mathcal{O}_{\Sigma}(\pmb{\sigma})^\top \mathcal{O}_{\Sigma}(\pmb{\sigma}))$ of each $\Sigma$ where $\Sigma$ varies through all possible SLSs.
The constant $c$ can be hence considered as a counterpart of VC dimensions in PAC-style
error bounds \cite{BOO:V99}, and it captures the complexity of the model class. 
Alternatively, $c$ could also be estimated from data
with high probability, as we will see in
Theorem \ref{thm!pract} in Section \ref{practical:issue}.

\begin{remark}
	In fact, our analysis in Section \ref{sec:probstability} can be
	extended to the case when the initial state is not uniformly distributed, but its distribution $\mathbb{P}$ satisfies the following regularity condition:
	There exists an increasing function $\nu: [0,1]\rightarrow [0,1]$ such that, for any $S\subseteq \mathbb{S}_{n-1}$, $\mathbb{P}\{x: x/\|x\| \in S\} \ge \nu(\mu(S))$, where $\mu(\cdot)$ is the uniform probability measure on the unit sphere $\mathbb{S}_{n-1}$. 
	This condition is satisfied for example for Gaussian distributions.
Under this condition, the inequality in \eqref{eqn:rhoSigQ} becomes
\begin{align}\label{eqn:rhoSigQnu}
	\rho(\Sigma) \le \frac{\gamma^*(\omega_N)}{\sqrt[T-k]{\delta\left(\frac{\nu^{-1}(\epsilon M^T)\chi_{\Sigma}(P^*(\omega_N),k)}{2}\right)}}.
\end{align}
\end{remark}

\begin{remark}
	The proposed approach can be also extended to disturbed systems with bounded additive disturbances in the form of 	
	\begin{equation}
		x(t+1) = A_{\sigma(t)}x(t) +w(t), ~ ~ ~ y(t) = C_{\sigma(t)}x(t) + v(t)
		\label{eq:LSSd}
	\end{equation}
where $w(t)$ and $v(t)$ are bounded. Similar probabilistic guarantees can be derived with a-priori information on the bounds of the disturbances.
\end{remark}

\section{Technical issues}
\label{practical:issue}
In this section, we discuss some technical issues of the proposed data-driven stability analysis approach.

\subsection{Sample complexity}\label{sec:samplex}
We first provide some insights into the relation between the size of the sample and the precision of the solution of the scenario program \eqref{eqn:scenariobreaking} from the perspective of PAC learning \cite{BOO:SB14}. Let us formally define the \emph{sample complexity} of the scenario program \eqref{eqn:scenariobreaking} below.
\begin{definition}[Sample complexity]\label{def:complexity}
	Given $\beta\in (0,1)$, $\varepsilon \ge 0$ and any integer $k\ge \mathcal{H}(\Sigma)$ , the sample complexity of the scenario program \eqref{eqn:scenariobreaking}, denoted by $N_s(\varepsilon,\beta)$, is the minimal $N\in \mathbb{Z}^+$ such that
	\begin{align}
		\mathbb{P}^N\{\omega_N: \gamma^o_k \le (1+\varepsilon)\gamma_k^*(\omega_N) \} \ge 1- \beta, 
	\end{align}
where $\gamma^o_k$ is given in \eqref{eqn:gammaoPo:alt}.
\end{definition}

In the following proposition,  we present some results on the \emph{sample complexity} for the proposed data-driven approach.
\begin{proposition}\label{prop:samplecomplexity}
	Suppose the conditions in Theorem \ref{thm:prostability} hold. For any $\beta\in (0,1)$ and $\varepsilon \ge 0$, let the sample complexity $N_s(\varepsilon,\beta)$ be defined as in Definition \ref{def:complexity}. Then, the following results hold:\\
	(i) There exists a constant $c\ge 1$ such that
	\begin{align}
		N_s(\varepsilon,\beta) &\le\mathcal{N}_{\phi}(\frac{\mathcal{I}(1-\frac{1}{(1+\varepsilon)^{2(T-k)}};\frac{n-1}{2},\frac{1}{2})}{M^T\chi(Q)\sqrt{(c\bar{\lambda})^{n-1}}},\beta) \label{eqn:Nvarepbetaphi} \\
		&=  O\left( \frac{1}{\beta} (\frac{1}{\varepsilon})^{\frac{n}{2}} \right). \label{eqn:Obetaepsilon}
	\end{align}
	where $\mathcal{N}_{\phi}(\epsilon,\beta)$ denotes the minimal $N\in \mathcal{Z}^+$ satisfying $\phi(\epsilon;d,N) \le \beta$ for any $\epsilon\in (0,1)$.\\
	(ii) When the system $\Sigma$ is not asymptotically stable under arbitrary switching, i.e., $\rho(\Sigma) > 1$, for any $N \ge N_s(\varepsilon,\beta)$,
	\begin{align}\label{eqn:gammainstablebeta}
		\mathbb{P}^N\{\omega_N: \gamma^*(\omega_N) < \frac{1}{1+\varepsilon} \} \le \beta.
	\end{align}
\end{proposition}

The proof is given in the appendix. From Proposition \ref{prop:samplecomplexity}, the \emph{sample complexity} is polynomial in $1/\varepsilon$ and $1/\beta$, which implies PAC learnability in the sense of Valiant's
definition in \cite{ART:V84}. In addition, Property (ii) of Proposition \ref{prop:samplecomplexity} provides a measure of the risk of stability learning using the proposed data-driven approach.
\begin{remark}
	Using tighter explicit bounds of $\phi(\epsilon;d,N)$ in \cite{ART:C10,ART:ATLR15}, it can be even shown that $N_s(\varepsilon,\beta) = O\left( \ln(\frac{1}{\beta}) (\frac{1}{\varepsilon})^{\frac{n}{2}}  \right).$
\end{remark}

Alternatively, we can also define sample complexity based on the absolute error, i.e.,  the minimal $N\in \mathbb{Z}^+$ such that
\begin{align}
	\mathbb{P}^N\{\omega_N: \gamma^o \le \gamma^*(\omega_N) + \varepsilon\} \ge 1- \beta.
\end{align}
The analysis above can be adapted to this alternative definition. To avoid repetition, we do not intend to provide the details.

\subsection{Pathwise observability index estimation}\label{sec:indexestimation}
The results in the previous section all rely on the condition that $k \ge \mathcal{H}(\Sigma)$.  However, the exact value of $\mathcal{H}(\Sigma)$ is often unavailable. Though explicit bounds on $\mathcal{H}(\Sigma)$ exist in \cite{babaali2003pathwise}, they are only useful for small values of $n$ and $M$. 

We now show an efficient procedure to estimate $\mathcal{H}(\Sigma)$. Let us  define, $\forall k \ge 1$,
\begin{subequations}\label{eqn:scenarioHo}
\begin{align}
	& \xi^o_k \coloneqq \inf_{\xi \ge 0} \xi\\
	& \textrm{s.t. } 
	Y_{\pmb{\sigma},x,k,k}^{\top}  Y_{\pmb{\sigma},x,k,k} \le
	\xi^{2k} Y_{\pmb{\sigma},x,0,k}^{\top}  Y_{\pmb{\sigma},x,0,k},
	\nonumber \\
	& 
	\qquad \forall (x,\pmb{\sigma})\in \mathbb{S}_{n-1}\times \mathcal{M}^{2k} 
\end{align}
\end{subequations}
where $Y_{\pmb{\sigma},x,k,k}$ and $Y_{\pmb{\sigma},x,0,k}$ are given in \eqref{eq:obs1}. The following proposition shows an important property of the sequence $\{\xi^o_k\}_{k\ge 1}$.
\begin{proposition}\label{prop:xio}
Suppose the \SLS\ $\Sigma$ is pathwise observable. Let  $\{\xi^o_k\}_{k\ge 1}$ be defined as in \eqref{eqn:scenarioHo}. Then,  $\xi^o_k < \infty $ if and only if $k \ge \mathcal{H}(\Sigma)$.
\end{proposition}
Given a sample set $\omega_N$ with a sufficiently long horizon $T$ and  the observed trajectories $\{y_t^i: 0\le t\le T-1, 1\le i \le N\}$, empirical estimates of $\{\xi^o_k: 1\le k\le \lfloor \frac{T}{2} \rfloor \}$ are given as
\begin{align}\label{eqn:scenarioH}
	\xi_k (\omega_N)\coloneqq  \max_{1\le i\le N} \frac{\|\tilde{v}_k^i\|}{\|v^i_k\|}, 
\end{align}
where $v^i_k$ is given in \eqref{eqn:vikzki} and 
\begin{align}
\tilde{v}_k^i  = \begin{pmatrix}
(y^i_k)^\top &
(y^i_{k+1})^\top &
\cdots &
(y^i_{2k-1})^\top
\end{pmatrix}^\top.	
\end{align}
With these estimates, we then consider the minimal $k$ such that $\xi_k (\omega_N)$ is less than some user-defined  threshold. This will be illustrated by a numerical example in the next section.


\subsection{Explicit bounds}\label{expl:bounds}
The probabilistic bound in Theorem \ref{thm:prostability} implicitly depends on the system matrices which are not available. In this section, we show that explicit bounds can also be derived by estimating $\chi_{\Sigma}(\cdot,k)$. 

Given any $k \ge \mathcal{H}(\Sigma)$ and a sample set $\omega_N$ with the corresponding measurements $\{y_t^i: t = 0,1,\cdots,T-1, i=1,2,\cdots, N\}$, let
\begin{align}
\overline{\zeta}_k(\omega_N) \coloneqq \max_{1\le i\le N} \|v^i_k\|, \quad  
\underline{\zeta}_k(\omega_N) \coloneqq \min_{1\le i\le N} \|v^i_k\|,
\label{eqn:underlinezeta}
\end{align}
where $v_k^i$ is defined in \eqref{eqn:vikzki}. For any $\varepsilon\in (0,1)$ and $\varepsilon'\in (0,1)$, let us define 
\begin{align}\label{eqn:psiomegaN}
\psi_{\varepsilon,\varepsilon'}(\omega_N) \coloneqq \frac{1}{\delta(\frac{\varepsilon M^k}{2}) \frac{\underline{\zeta}_k(\omega_N)}{\overline{\zeta}_k(\omega_N)} -   \sqrt{2-2\delta(\frac{\varepsilon' M^k}{2})} } .
\end{align}
This allows to estimate $\chi_{\Sigma}(\cdot,k)$ in Theorem \ref{thm:prostability}, which leads to 
an explicit probabilistic upper bound on $\rho(\Sigma)$, as stated in the following theorem.
\begin{theorem}
\label{thm!pract}
Consider the same conditions as in Theorem \ref{thm:prostability}. 
For any $\bar{\epsilon},\varepsilon, \varepsilon' \in (0,1)$, with probability no smaller than $1-\phi(\bar{\epsilon};d,N)-(1-\varepsilon)^N-(1-\varepsilon')^N$, 
\begin{align}
	\rho(\Sigma) \le  \frac{\gamma^*(\omega_N)}{\sqrt[T-k]{\delta\left(\frac{\bar{\epsilon} M^T \left(\psi_{\varepsilon,\varepsilon'}(\omega_N) \right)^{n-1}\sqrt{\left(P^*(\omega_N)\right)^{n-1}}}{2}\right)}}
\end{align}
where $\psi_{\varepsilon,\varepsilon'}(\omega_N)$ is given in \eqref{eqn:psiomegaN}.
\end{theorem}

\begin{remark}[Checking stability]
	\label{rem:stab_check}
	We are now ready to continue our discussion on checking stability. As it was
	mentioned in Section \ref{sec:leaning:estim}, the main idea is to solve \eqref{eqn:scenariobreaking} and then to check if
	$\gamma^*_k(\omega_N)$ is sufficiently smaller than $1$. If we want to make this idea reliable,
	we can proceed as follows. One option is to choose the confidence
	level $\beta$ and check if the right-hand side of \eqref{eqn:jsrapriori}
	is smaller than $1$, i.e., if  $\gamma^*_k(\omega_N) \le \sqrt[T-k]{\delta\left(\frac{\epsilon M^T\sqrt{(c\bar{\lambda})^{n-1}}}{2}\right)}$.
	We can then conclude that the underlying system is stable, with probability $1-\beta$
	over the data. The drawback of this approach that we need to estimate $c$, for
	which we need some a-priori knoweldge on the underlying system.
	Another option is to use Theorem \ref{thm!pract}: we choose 
	$\epsilon,\epsilon^{'}$ such that 
	$\beta=\phi(\bar{\epsilon};d,N)-(1-\varepsilon)^N-(1-\varepsilon')^N$ and then 
	we check if $\gamma^*_k(\omega_N) \le \sqrt[T-k]{\delta\left(\frac{\bar{\epsilon} M^T \left(\psi_{\varepsilon,\varepsilon'}(\omega_N) \right)^{n-1}\sqrt{\left(P^*(\omega_N)\right)^{n-1}}}{2}\right)}$.
	If the latter is the case, then the underlying system is stable with probability
	$1-\beta$ over the sampled data. 
	Note that $P^*(\omega_N)$ is computed at the same time as $\gamma^*_k(\omega_N)$, when
	\eqref{eqn:scenariobreaking} is solved. 
\end{remark}

\section{Numerical examples}
\label{sec:num}
In this section, we illustrate the performance of the proposed approach on numerical examples. To show the advantages of our approach, we make a comparison with the identification-based approach in which we identify the model using hybrid system identification techniques and analyze the stability based on the model. Our codes are available via \url{https://github.com/zhemingwang/DataDrivenStabilityAnalysis}, and
all experiments are run on a MacBook Pro with an Intel Core i7 and 16 GB of RAM.

\subsection{Comparison with hybrid system identification} 
To analyze properties of the system, a natural idea is to first identify a dynamical model which then allows us to use  well-documented  model-based techniques. For hybrid systems, there also exist many identification techniques, see, e.g., \cite{IdentSurvey,BOO:LB18}.  In the comparison, we consider switched auto-regressive models, which are quite popular in hybrid system identification, see, e.g.,  \cite{VidalAutomatica}. For stability analysis, we convert the switched auto-regressive model into its state-space form using the transformation described in \cite{weiland2006equivalence} and compute the JSR of the augmented model using  the JSR toolbox \cite{INP:VHJ14}. Note that the equivalence between switched affine models and switched ARX models has been proved in \cite{weiland2006equivalence} under the pathwise observability condition. In particular, let us emphasize that such a conversion preserves internal stability under the pathwise observability condition. The key step in the identification procedure is clustering where we group or label the output trajectories. For each group, we then estimate the parameters of its auto-regressive model by solving a least-square problem.  A few clustering algorithms that we use in the simulation are the following: Generalized Principal Component Analysis (GPCA) \cite{vidal2005generalized}, Sparse Subspace Clustering (SSC) \cite{elhamifar2013sparse}, Sparse Subspace Clustering by Orthogonal Matching Pursuit (SSC-OMP), \cite{you2016scalable}, Elastic net Subspace Clustering (EnSC) \cite{you2016oracle}, and Piecewise Affine Regression and Classification (PARC) \cite{bemporad2022piecewise}.

\begin{figure}[h]
				\centering
				\includegraphics[width=0.9\linewidth]{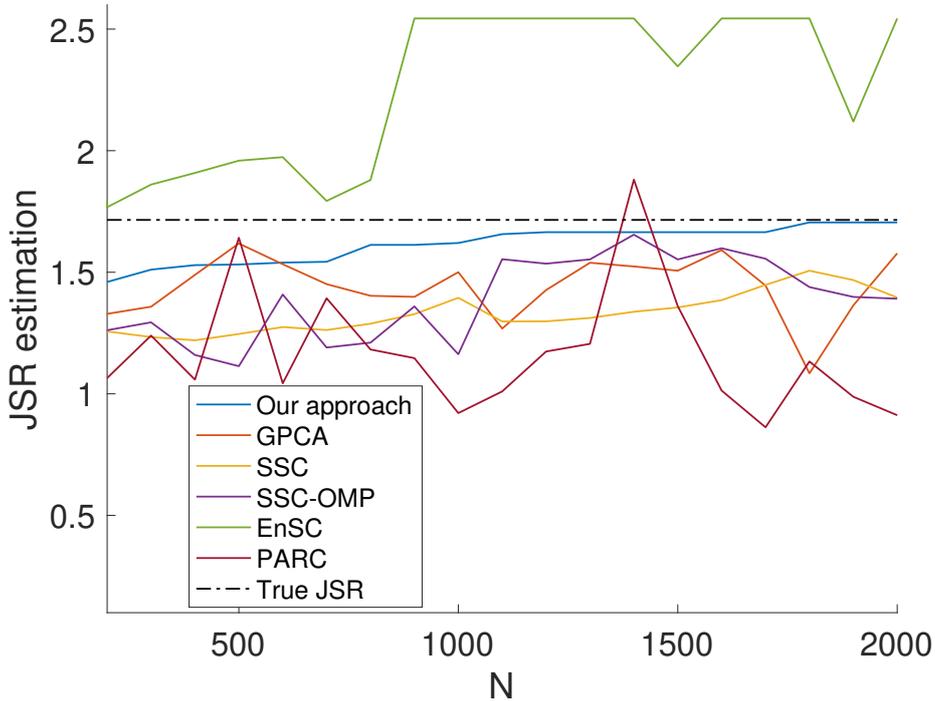}
				\caption{Comparison with the identification-based approach using different hybrid system identification algorithms.}
				\label{fig:comp}
			\end{figure}

The numerical example is a three-dimensional switched linear system with $3$ modes and $2$ outputs.
			We then generate an output data set of length $T=5$ and set $k=3$ in \eqref{eqn:vikzki}. The order of the auto-regressive models is  set to be $k$ and the maximal number of groups in the clustering algorithms is set to be $M^{k-1}$. The results are given in Figure \ref{fig:comp}, which shows the JSR estimation of our approach and the identification-based approach for different clustering algorithms as the number of data points increases. From this figure, we can see that the modeling error in hybrid system identification can lead to a false stability inference. Note that more data does not necessarily result in more accurate system identification from an algorithmic point of view, especially when there is a clustering step in the whole procedure. This is supported by Figure \ref{fig:comp} as the JSR estimation can become worse as $N$ increases. In contrast, our approach is guaranteed to provide better JSR estimation as $N$ increases. Another practical issue of the identification-based approach is that in the clustering algorithms there are some parameters that need to be tuned carefully. We also make a comparison with the two standard clustering algorithms GPCA and SSC in terms of computational time in Table \ref{tab:time}.
			\begin{table}[h]
				\renewcommand{\arraystretch}{1.2}
				\centering
				\begin{tabular}{|c|c|c|c|c|c|} 
					\hline
					\multicolumn{2}{|c|}{$N$}   & $500$  & $1000$ & $1500$  & $2000$   \\ 
					\hline
					\multicolumn{2}{|c|}{Our approach}   & $22.08$  & $50.53$ & $61.18$ &  $62.48$ \\ 
					\hline
					\multirow{2}{*}{GPCA} & $T_{id}$   & $4.92$  & $40.46$  & $137.90$ & $256.27$  \\ 
					\cline{2-6}
					&  $T_{JSR}$   & $15.42$  & $32.93$  & $14.22$  &  $5.19$ \\ 
					\hline
					\multirow{2}{*}{SSC} &  $T_{id}$  & $894.20$  & $2550.08$ & $3554.40$  & $5015.54$  \\ 
					\cline{2-6}
					& $T_{JSR}$      & $67.90$  & $124.18$ & $67.10$ & $120.97$  \\
					\hline
				\end{tabular}
				\caption{Computational time (second) for different algorithms:  $T_{id}$ denotes the computational time for the identification step and $T_{JSR}$ denotes the computational time for computing the JSR using the model.}
				\label{tab:time}
			\end{table}
			
			\noindent
			\subsection{A six-dimensional example} 
\begin{figure}[h]
				\centering
				\includegraphics[width=0.8\linewidth]{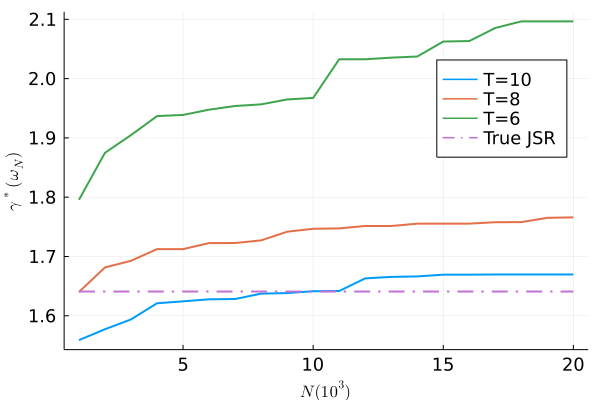}
				\caption{Data-based convergence rates for different horizon lengths.}
				\label{fig:gamma}
			\end{figure}

			We also apply the proposed data-driven approach to  a higher-dimensional example. We consider a switched linear system with $n=6,M=3,p=2$ where the entries of the dynamics and output matrices are chosen randomly from the uniform distribution over $[-1,1]$. 
			We then sample a finite set of initial states following the Gaussian distribution $\mathcal{N}(0,I)$ and project them onto the unit sphere. In this way, we obtain a set of initial states that are uniformly distributed on the unit sphere. With this data set, we formulate Problem \eqref{eqn:scenariobreaking} with $k=3$. This is a valid option as it can be verified that the system is \emph{pathwise observable} with the \emph{pathwise observability index} $\mathcal{H}(\Sigma)=2$. When the information of $\mathcal{H}(\Sigma)$ is not available, we use the procedure in Section \ref{sec:indexestimation} to estimate $\mathcal{H}(\Sigma)$. Finally, we are ready to solve Problem \eqref{eqn:scenariobreaking}. The simulation results are given in Figure \ref{fig:gamma} for different values of horizon length $T$. As expected, $\gamma^*(\omega_N)$ gradually approaches the true JSR (which can be obtained from the JSR toolbox \cite{INP:VHJ14}) as $N$ and $T$ increase. 
			\begin{figure}[h]
				\centering
				\includegraphics[width=0.8\linewidth]{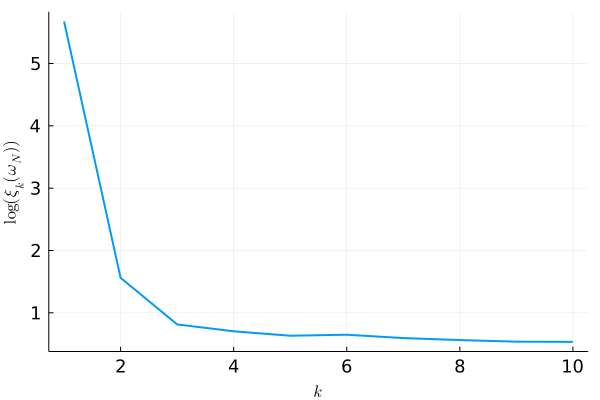}
				\caption{Estimates of $\xi_k^o$ for different values of $k$.}
				\label{fig:xik}
			\end{figure}

			
			%
			%

			
			\section{Conclusions}
			In this paper, we have studied the problem of output-based stability learning of a special family of hybrid systems, namely switched linear systems, using a time-series output data set. We leverage a recently introduced approach for state-based data-driven stability analysis. However, the fact that only a (possibly low-dimensional) output is observed incurs additional theoretical challenges that we tackle here. With the proposed output-based Lyapunov framework, we are able to derive probabilistic stability certificates for partially observed switched linear systems. We have also presented additional procedures which allow to use explicit upper bounds on the JSR of switched linear systems. To validate our approach, we have made a comparison with the identification-based approach in which a model is identified using existing clustering algorithms in the literature. Numerical results suggest that it is beneficial to use our direct analysis approach as modeling errors can lead to false stability inferences.

			\appendix
			\setcounter{secnumdepth}{0}
			\section{Appendix}
			\setcounter{secnumdepth}{1}
			\renewcommand{\thesection}{A\arabic{section}}
			
			\section{Proof of Theorem \ref{thm:jsrS}}\label{sec:almostproof}
			Since $\Sigma$ is pathwise observable, $\mathcal{P}^{\pmb{\sigma}}_\Sigma(P,0,k)\succ 0$ and $\mathcal{P}^{\pmb{\sigma}}_\Sigma(P,\ell,k) \succ 0$ for any $\pmb{\sigma} \in \mathcal{M}^{k+\ell}$. We pick an arbitrary $\pmb{\sigma} \in \mathcal{M}^{k+\ell}$ and consider the Cholesky decompositions of $\mathcal{P}^{\pmb{\sigma}}_\Sigma(P,0,k)$ and $\mathcal{P}^{\pmb{\sigma}}_\Sigma(P,\ell,k)$:
			\begin{align*}
				U^\top U =\mathcal{P}^{\pmb{\sigma}}_\Sigma(P,0,k), \tilde{U}^\top \tilde{U} =\mathcal{P}^{\pmb{\sigma}}_\Sigma(P,\ell,k),
			\end{align*}
			where $U \in \mathbb{R}^{n\times n}$ and $\tilde{U} \in \mathbb{R}^{n\times n}$ are upper triangular matrices.  Hence, by the change of coordinates, \eqref{eqn:AOPgammax} becomes
			\begin{align}
				&(\tilde{U}A_{\pmb{\sigma}_{0:\ell-1}} U^{-1}x)^\top \tilde{U}A_{\pmb{\sigma}_{0:\ell-1}} U^{-1}x \nonumber\\
				\le &\gamma^{2\ell} x^\top x, \quad \forall x\in U(\mathbb{S}_{n-1}\setminus S).
			\end{align}
			The homogeneity of the dynamics implies that
			\begin{align}\label{eqn:Sproject}
				&(\tilde{U}A_{\pmb{\sigma}_{0:\ell-1}} U^{-1}x)^\top \tilde{U}A_{\pmb{\sigma}_{0:\ell-1}} U^{-1}x \nonumber\\
				\le &\gamma^{2\ell} x^\top x,  \quad \forall x\in \Pi_{\mathbb{S}_{n-1}}\left(U(\mathbb{S}_{n-1}\setminus S)\right),
			\end{align}
			where $\Pi_{\mathbb{S}_{n-1}}(\cdot)$ denote the projection onto $\mathbb{S}_{n-1}$. Equivalently, we can write \eqref{eqn:Sproject} as
			\begin{align}\label{eqn:Sprojecteq}
				&(\tilde{U}A_{\pmb{\sigma}_{0:\ell-1}} U^{-1}x)^\top \tilde{U}A_{\pmb{\sigma}_{0:\ell-1}} U^{-1}x \nonumber\\
				\le &\gamma^{2\ell} x^\top x,  \quad \forall x\in \mathbb{S}_{n-1} \setminus \Pi_{\mathbb{S}_{n-1}}\left(U S\right).
			\end{align}
			Following the same arguments in Theorem 15 in \cite{ART:KBJT19}, we obtain that
			\begin{align}
				\tilde{U}A_{\pmb{\sigma}_{0:\ell-1}} U^{-1} \mathbb{S}_{n-1} \subset \frac{\gamma^\ell}{\delta\left( \frac{\mu(S) \chi(\mathcal{P}^{\pmb{\sigma}}_\Sigma(P,0,k)) }{2} \right)} \mathbb{B}_n.
			\end{align}
			An illustration of the function $\delta(\cdot)$  is given in Figure \ref{fig:delta}. Hence, it holds that
			\begin{align}
				&\left( \tilde{U}A_{\pmb{\sigma}_{0:\ell-1}} U^{-1}  \right)^\top \tilde{U}A_{\pmb{\sigma}_{0:\ell-1}} U^{-1} \nonumber\\
				\preceq  &\frac{\gamma^{2\ell}}{ (\delta\left( \frac{\mu(S) \chi(\mathcal{P}^{\pmb{\sigma}}_\Sigma(P,0,k)) }{2} \right))^2} I,
			\end{align}
			which implies that
			\begin{align}
				&A_{\pmb{\sigma}_{0:\ell-1}}^\top \mathcal{P}^{\pmb{\sigma}}_\Sigma(P,\ell,k)A_{\pmb{\sigma}_{0:\ell-1}}  \nonumber \\
				\preceq  &\frac{\gamma^{2\ell}}{ (\delta\left( \frac{\mu(S) \chi(\mathcal{P}^{\pmb{\sigma}}_\Sigma(P,0,k)) }{2} \right))^2} \mathcal{P}^{\pmb{\sigma}}_\Sigma(P,0,k). 
			\end{align}
			As $\pmb{\sigma}$ is chosen arbitrarily, the inequality above holds for any $\pmb{\sigma} \in \mathcal{M}^{k+\ell}$. Note that the function $\delta(\cdot)$ is non-increasing. Therefore, with the definition of $\chi_{\Sigma}(P,k)$ as in \eqref{eqn:kappaSigma}, we arrive at
			\begin{align}
				&A_{\pmb{\sigma}_{0:\ell-1}}^\top \mathcal{P}^{\pmb{\sigma}}_\Sigma(P,\ell,k)A_{\pmb{\sigma}_{0:\ell-1}}  \nonumber \\
				\preceq  &\frac{\gamma^{2\ell}}{ (\delta\left( \frac{\mu(S) \chi_{\Sigma}(P,k) }{2} \right))^2} \mathcal{P}^{\pmb{\sigma}}_\Sigma(P,0,k), \forall \pmb{\sigma} \in \mathcal{M}^{k+\ell}.
			\end{align}
			Finally, from Theorem \ref{thm:jsr}, we arrive at \eqref{eqn:jsrS}.

			\section{Proof Theorem \ref{thm:prostability}}
			
			

			To simplify the notation, we drop the subscript $k$ in $(\gamma_k^*(\omega_N),P_k^*(\omega_N))$ in the proof. We use the scenario approach to derive a probabilistic upper bound on the JSR via geometric analysis, following \cite{ART:KBJT19,INP:BJW21}. From Theorem 6 in \cite{INP:BJW21}, the chance-constrained theorem in \cite{ART:CG08,ART:C10} is applicable and we obtain that for any $\epsilon\in (0,1)$,
			\begin{align*}
			\mathbb{P}^N\{\omega_N: \mathbb{P} \{V(\omega_N)\} > \epsilon \} \le \phi(\epsilon;d,N), 
			\end{align*}
			where 
			\begin{align*}
				V(\omega_N)\coloneqq &  \{ (x,\pmb{\sigma}) \in\mathbb{S}_{n-1}\times \mathcal{M}^{T}: \\
				&x^\top A_{\pmb{\sigma}_{0:\ell-1}}^\top \mathcal{P}^{\pmb{\sigma}}_{\Sigma}(P^*(\omega_N)),T-k,k)A_{\pmb{\sigma}_{0:\ell-1}}x\\
				>& (\gamma^*(\omega_N))^{2(T-k)} x^\top \mathcal{P}^{\pmb{\sigma}}_{\Sigma}(P^*(\omega_N),0,k)x \}.
			\end{align*}
			For any $\pmb{\sigma} \in \mathcal{M}^{T}$, let $V^{\pmb{\sigma}}(\omega_N)\coloneqq \{x \in \mathbb{S}_{n-1}: (x,\pmb{\sigma}) \in V(\omega_N)\}$. With this definition, we claim that, for any $\pmb{\sigma} \in \mathcal{M}^{T}$
			$x\in \mathbb{S}_{n-1}\setminus \bigcup_{\pmb{\sigma}\in \mathcal{M}^T} V^{\pmb{\sigma}}(\omega_N)$
			\begin{align*}
				&x^\top A_{\pmb{\sigma}_{0:\ell-1}}^\top \mathcal{P}^{\pmb{\sigma}}_{\Sigma}(P^*(\omega_N)),T-k,k)A_{\pmb{\sigma}_{0:\ell-1}} x \\
				\le  & (\gamma^*(\omega_N))^{2(T-k)} x^\top \mathcal{P}^{\pmb{\sigma}}_{\Sigma}(P^*(\omega_N),0,k)x
			\end{align*}
			Note that in the worst case, the sets $\{V^{\pmb{\sigma}}(\omega_N)\}$  are disjoint.
			When $\mathbb{P} \{V(\omega_N)\} \le \epsilon$,  $\mu(\bigcup_{\pmb{\sigma}\in \mathcal{M}^T} V^{\pmb{\sigma}}(\omega_N)) \le \epsilon M^T$, where $\mu(\cdot)$ denotes the (probability) uniform measure on $\mathbb{S}_{n-1}$. From Theorem \ref{thm:jsrS}, we conclude that, with probability no smaller than $1-\phi(\epsilon;d,N)$, \eqref{eqn:rhoSigQ} holds.

			\section{Proof of Proposition \ref{prop:samplecomplexity}}
			Proof: (i) Given $\beta\in (0,1)$ and $\varepsilon >0$, from Corollary \ref{cor:complexity}, we look for a $N$ such that $\eta(\beta,N) \le 1+\varepsilon$, which is equivalent to 
			\begin{align}\label{eqn:IxabK2}
				\phi(\frac{\mathcal{I}(1-\frac{1}{(1+\varepsilon)^{2(T-k)}};\frac{n-1}{2},\frac{1}{2})}{K_2},d,N) \le \beta
			\end{align}
			where $K_2 = M^T\chi(Q)\sqrt{(c\bar{\lambda})^{n-1}}$. Hence, \eqref{eqn:Nvarepbetaphi} holds. We then show that $\mathcal{N}(\varepsilon,\beta)=O\left( \frac{1}{\beta} \frac{1}{\varepsilon^{\frac{n}{2}}}  \right)$ as $\varepsilon \rightarrow 0$. To do so, we first show that $\phi(\epsilon;d,N)$ can be explicitly bounded from above by $\frac{d}{\epsilon(N+1)}$ for any $\epsilon\in (0,1)$ and any $N\in \mathbb{Z}^+$. This can be verified using the following manipulations:
			\begin{align}
				\frac{\phi(\epsilon;d,N)}{\frac{d}{\epsilon(N+1)}} &= 	\frac{\sum_{i=0}^{d-1} {N \choose i} \epsilon^i(1-\epsilon)^{N-i}}{\frac{d}{\epsilon(N+1)}} \nonumber\\
				& = \sum_{i=0}^{d-1} \frac{i+1}{d}{N+1 \choose i+1} \epsilon^{i+1}(1-\epsilon)^{N-i} \nonumber\\
				&\le  \sum_{i=1}^{d} {N+1 \choose i} \epsilon^{i}(1-\epsilon)^{N+1-i} \nonumber\\
				&\le \phi(\epsilon;d+1,N+1) \le 1. \label{eqn:phibound}
			\end{align}
			This means that 
			\begin{align}\label{eqn:Nphibetaepsilon}
				\mathcal{N}_{\phi}(\epsilon,\beta) \le \frac{d}{\beta \epsilon}.
			\end{align}
			We then show that $\mathcal{I}(x;\frac{n-1}{2},\frac{1}{2}) \ge K_1 x^{\frac{n}{2}}$, where $\mathcal{I}(x;a,b)$ is defined as in \eqref{eqn:Ixab} and $K_1=\frac{\frac{4}{n}}{\int_0^1 t^{\frac{n-3}{2}}(1-t)^{-\frac{1}{2}}dt}$, as follows
			\begin{align}
				\mathcal{I}(x;\frac{n-1}{2},\frac{1}{2})  &=  \frac{\int_0^x t^{\frac{n-3}{2}}(1-t)^{-\frac{1}{2}}dt}{\int_0^1 t^{\frac{n-3}{2}}(1-t)^{-\frac{1}{2}}dt}  \nonumber\\
				&=  \frac{\int_0^x t^{\frac{n-2}{2}} \frac{1}{(t(1-t))^{\frac{1}{2}}}dt}{\int_0^1 t^{\frac{n-3}{2}}(1-t)^{-\frac{1}{2}}dt} \nonumber\\
				& \ge \frac{2\int_0^x t^{\frac{n-2}{2}} dt}{\int_0^1 t^{\frac{n-3}{2}}(1-t)^{-\frac{1}{2}}dt} = K_1 x^{\frac{n}{2}}, \label{eqn:Ixabbound}
			\end{align}
			where the inequality is due to the fact that $t(1-t) \le \frac{1}{4}$.  Combing \eqref{eqn:Nphibetaepsilon} and \eqref{eqn:Ixabbound} yields
			\begin{align*}
				&\mathcal{N}_{\phi}(\frac{\mathcal{I}(1-\frac{1}{(1+\varepsilon)^{2(T-k)}};\frac{n-1}{2},\frac{1}{2})}{M^T\chi(Q)\sqrt{(c\bar{\lambda})^{n-1}}},\beta) \\
				\le & \frac{K_2d}{\beta \mathcal{I}(1-\frac{1}{(1+\varepsilon)^{2(T-k)}};\frac{n-1}{2},\frac{1}{2})}\\
				\le & \frac{K_2d}{\beta K_1(1-\frac{1}{(1+\varepsilon)^{2(T-k)}})^{\frac{n}{2}}}\\
				= & \frac{K_2 d}{K_1 \beta} \frac{(1+\frac{1}{\varepsilon})^{n(T-k)}}{\left((1+\frac{1}{\varepsilon})^{2(T-k)}-(\frac{1}{\varepsilon})^{2(T-k)}\right)^{\frac{n}{2}}}.
			\end{align*}
			By some manipulations, we get that $N_s(\varepsilon,\beta) \le \frac{K_2 d}{K_1 \beta} (\frac{1}{\varepsilon})^{\frac{n}{2}}$ as $\varepsilon \rightarrow 0$, which can be equivalently expressed as \eqref{eqn:Obetaepsilon}. This completes the proof. \\
			(ii) From the definition of sample complexity in Definition \ref{def:complexity}, it holds that $\mathbb{P}^N\{\omega_N: \gamma^o > (1+\varepsilon)\gamma^*(\omega_N) \}  \le \beta$ for any $N \ge N_s(\varepsilon,\beta)$. The fact that $\rho(\Sigma) > 1$ implies that $\gamma^o >1$. With this, it can be verified that $\{\omega_N: \gamma^*(\omega_N) < \frac{1}{1+\varepsilon}\} \subseteq \{\omega_N: \gamma^o > (1+\varepsilon)\gamma^*(\omega_N)\}$, which implies \eqref{eqn:gammainstablebeta}.

			\section{Proof of Proposition \ref{prop:xio}}
			When $k \ge \mathcal{H}(\Sigma)$, $\textrm{rank}\left(\mathcal{O}_{\Sigma}(\pmb{\sigma}) \right) = n$ for any $\pmb{\sigma} \in \mathcal{M}^{k}$. Hence, $\xi^o_k$ is bounded. (Necessity) The proof goes by contradiction. Suppose there exists $k<\mathcal{H}(\Sigma)$ such that $\xi^o_k < \infty $.  From the definition of $\mathcal{H}(\Sigma)$, there exists at least one sequence $\tilde{\pmb{\sigma}} \in \mathcal{M}^{k}$ such that $\textrm{rank}\left(\mathcal{O}_{\Sigma}(\tilde{\pmb{\sigma}}) \right) < n$. The fact that $\xi^o_k < \infty $ implies that $\mathcal{O}_{\Sigma}(\tilde{\pmb{\sigma}})x = 0$ implies $\mathcal{O}_{\Sigma}(\pmb{\sigma}) A_{\tilde{\pmb{\sigma}}}x=0$ for any $\pmb{\sigma} \in  \mathcal{M}^{k}$. In particular, $\mathcal{O}_{\Sigma}(\tilde{\pmb{\sigma}})x = 0$ implies $\mathcal{O}_{\Sigma}(\tilde{\pmb{\sigma}}) A_{\tilde{\pmb{\sigma}}}x=0$. Then, there exists a matrix $G\in \mathbb{R}^{pk\times pk}$ such that $\mathcal{O}_{\Sigma}(\tilde{\pmb{\sigma}}) A_{\tilde{\pmb{\sigma}}} = G\mathcal{O}_{\Sigma}(\tilde{\pmb{\sigma}})$, i.e., each row of $\mathcal{O}_{\Sigma}(\tilde{\pmb{\sigma}}) A_{\tilde{\pmb{\sigma}}}$ can be written as a linear combination of the rows in $\mathcal{O}_{\Sigma}(\tilde{\pmb{\sigma}})$. As a result, the periodic sequence $\tilde{\pmb{\sigma}} \tilde{\pmb{\sigma}} \cdots$ is  unobservable, which contradicts the fact that $\Sigma$ is pathwise observable.

			\section{Proof of Theorem \ref{thm!pract}}
			To prove Theorem \ref{thm!pract}, we first state the following bound
			on the singular value of the observability matrices.
			\begin{lemma}\label{lem:normbound}
				Consider the same conditions as in Theorem \ref{thm:prostability}.
				Then,  for any $\varepsilon\in (0,1)$ and $\varepsilon'\in (0,1)$, with probability no smaller than 
				$1-(1-\varepsilon)^N-(1-\varepsilon')^N$, 
				\begin{align}\label{eqn:sigmamaxminpsi}
					\frac{\max_{\pmb{\sigma}\in \mathcal{M}^k}\sigma_{\max}(\mathcal{O}_{\Sigma}(\pmb{\sigma}))}{\min_{\pmb{\sigma}\in \mathcal{M}^k}\sigma_{\min}(\mathcal{O}_{\Sigma}(\pmb{\sigma})) } \le \psi(\omega_N) 
				\end{align}
			\end{lemma}
			Proof: Since only the first $k$ steps of the trajectory are relevant, we only consider the sample set $\omega_N$ of length $k$. The proof consists of three steps:\\
			Step 1: We consider the robust optimization problem below:
			\begin{align}\label{eqn:zetasigmamax}
				\min_{\zeta \ge 0} \zeta: \|\mathcal{O}_{\Sigma}(\pmb{\sigma})x\| \le \zeta, \forall (x,\pmb{\sigma})\in \mathbb{S}_{n-1}\times \mathcal{M}^k.
			\end{align}
			The optimum is exactly $\max_{\pmb{\sigma}\in \mathcal{M}^k}\sigma_{\max}(\mathcal{O}_{\Sigma}(\pmb{\sigma}))$. We want to show that the solution $\overline{\zeta}_k(\omega_N)$ in \eqref{eqn:underlinezeta} provides a probabilistic upper bound for Problem \eqref{eqn:zetasigmamax}. To do so, we show the following chance-constrained result, given an $\varepsilon \in (0,1)$,
			\begin{align}\label{eqn:omegaNscalar}
				\mathbb{P}^N \{ \omega_N:  \Omega(\overline{\zeta}_k(\omega_N))   > \varepsilon\} = (1-\varepsilon)^N
			\end{align}
			where $\Omega(\zeta) \coloneqq \mathbb{P}\{(x,\pmb{\sigma})\in \mathbb{S}_{n-1}\times \mathcal{M}^k: \|\mathcal{O}_{\Sigma}(\pmb{\sigma})x\| >\zeta \}$. The function $\Omega: [0,\max_{\pmb{\sigma}\in \mathcal{M}^k}\sigma_{\max}(\mathcal{O}_{\Sigma}(\pmb{\sigma})) ) \rightarrow [0,1]$ is decreasing. Hence, there exists a unique $\zeta$ such that $\Omega(\zeta) = \varepsilon$, denoted by $\zeta_{\varepsilon}$. With this, the set $\{\omega_N:  \Omega(\overline{\zeta}_k(\omega_N))   > \varepsilon\}$ can be equivalently expressed as $\{\omega_N: \omega_N \cap \{(x,\pmb{\sigma}): \|\mathcal{O}_{\Sigma}(\pmb{\sigma})x\| \ge \zeta_{\varepsilon} \} = \emptyset \}$ (whose measure is $(1-\varepsilon)^N$), which leads to \eqref{eqn:omegaNscalar}. Let us define the projected violating subset $\tilde{\mathbb{S}}$ as follows:
			\begin{align*}
				\tilde{\mathbb{S}} \coloneqq \{x\in \mathbb{S}_{n-1}: \exists \pmb{\sigma} \in \mathcal{M}^k, \|\mathcal{O}_{\Sigma}(\pmb{\sigma})x\| > \overline{\zeta}_k(\omega_N) \}.
			\end{align*}
			For any $\pmb{\sigma} \in \mathcal{M}^k$, we also define:
			\begin{align}
				\tilde{\mathbb{S}}_{\pmb{\sigma}} \coloneqq \{x\in \mathbb{S}_{n-1}:  \|\mathcal{O}_{\Sigma}(\pmb{\sigma})x\| > \overline{\zeta}_k(\omega_N) \}
			\end{align}
			By definition, $\tilde{\mathbb{S}} = \cup_{\pmb{\sigma} \in \mathcal{M}^k} \tilde{\mathbb{S}}_{\pmb{\sigma}}$, which implies that
			\begin{align}
				\mathbb{P}_x\{\tilde{\mathbb{S}}\} \le  \sum_{\pmb{\sigma} \in \mathcal{M}^k}\mathbb{P}_x\{\tilde{\mathbb{S}}_{\pmb{\sigma}}\}
			\end{align}
			where $\mathbb{P}_x$ denotes the uniform probability measure on $\mathbb{S}_{n-1}$ and the equality holds when the sets $\{\tilde{\mathbb{S}}_{\pmb{\sigma}}\}_{\pmb{\sigma} \in \mathcal{M}^k}$ are disjoint. With the inequality above, we get that
			\begin{align}
				\Omega(\overline{\zeta}_k(\omega_N)) &= \sum_{\pmb{\sigma} \in \mathcal{M}^k}\mathbb{P}_x\{\tilde{\mathbb{S}}_{\pmb{\sigma}}\} \mathbb{P}_{\sigma} \{\pmb{\sigma}\} = \frac{1}{M^k} \sum_{\pmb{\sigma} \in \mathcal{M}^k}\mathbb{P}_x\{\tilde{\mathbb{S}}_{\pmb{\sigma}}\} \nonumber\\
				& \ge  \frac{1}{M^k}\mathbb{P}_x\{\tilde{\mathbb{S}}\}
			\end{align}
			where $\mathbb{P}_{\sigma}$ denote the uniform distribution on $\mathcal{M}^k$. This means that $\Omega(\overline{\zeta}_k(\omega_N)) \le \varepsilon$ implies $\mathbb{P}_x\{\tilde{\mathbb{S}}\} \le M^k\varepsilon$. Hence, 
			\begin{align}
				\mathbb{P}^N \{ \omega_N:  \mathbb{P}_x\{\tilde{\mathbb{S}}\} \le \varepsilon M^k \} = (1-\varepsilon)^N
			\end{align}
			from \eqref{eqn:omegaNscalar}. Finally, following the same lines as the proof of \cite[Theorem 15]{ART:KBJT19}, we conclude that, with probability no smaller than $1-(1-\varepsilon)^N$, 
			\begin{align}\label{eqn:maxsigmazetak}
				\max_{\pmb{\sigma}\in \mathcal{M}^k}\sigma_{\max}(\mathcal{O}_{\Sigma}(\pmb{\sigma})) \le \frac{\overline{\zeta}_k(\omega_N)}{\delta(\frac{\varepsilon M^k}{2})}.
			\end{align}
			\\
			Step 2: Similarly, we define the following robust optimization problem:
			\begin{align}\label{eqn:zetasigmamin}
				\max_{\zeta \ge 0} \zeta: \|\mathcal{O}_{\Sigma}(\pmb{\sigma})x\| \ge \zeta, \forall (x,\pmb{\sigma})\in \mathbb{S}_{n-1}\times \mathcal{M}^k.
			\end{align}
			As the constraint above is not convex in $x$, we cannot repeat the same reasoning in (i). Nevertheless, it still holds that the optimum of \eqref{eqn:zetasigmamin} is $\min_{\pmb{\sigma}\in \mathcal{M}^k}\sigma_{\min}(\mathcal{O}_{\Sigma}(\pmb{\sigma}))$, which can be attained, i.e., there exists $(x^*,\pmb{\sigma}^*)$ such that $$\|\mathcal{O}_{\Sigma}(\pmb{\sigma}^*)x^*\|=\min_{\pmb{\sigma}\in \mathcal{M}^k}\sigma_{\min}(\mathcal{O}_{\Sigma}(\pmb{\sigma})).$$ For any $\varepsilon \in (0,1)$, we define the set $\bar{\mathbb{S}} \coloneqq \{x\in \mathbb{S}_{n-1}: |x^\top x^*| \ge \delta(\frac{\varepsilon M^k}{2})\}$ with $\mathbb{P}_x\{\bar{\mathbb{S}}\} = \varepsilon M^k$. The probability that $\omega_N \cap \bar{\mathbb{S}} \times \{\pmb{\sigma}^*\} \not= \emptyset$ is $1-(1-\varepsilon)^N$. In this case, there exists $(\bar{x},\bar{\pmb{\sigma}}) \in \omega_N$ such that $\pmb{\sigma}=\pmb{\sigma}^*$ and $|\bar{x}^\top x^*| \ge \delta(\frac{\varepsilon M^k}{2})$, which implies that $\|\bar{x}-x^*\| \le \sqrt{2-2\delta(\frac{\varepsilon M^k}{2})}$ or $\|\bar{x}+x^*\| \le \sqrt{2-2\delta(\frac{\varepsilon M^k}{2})}$. From the definition in \eqref{eqn:underlinezeta}, $ \|\mathcal{O}_{\Sigma}(\bar{\pmb{\sigma}})\bar{x}\| \ge \underline{\zeta}_k(\omega_N)$. We then consider the case that $\|\bar{x}-x^*\| \le \sqrt{2-2\delta(\frac{\varepsilon M^k}{2})}$ (the analysis is exactly the same for the other case). With these relations, it holds that
			\begin{align*}
				\|\mathcal{O}_{\Sigma}(\pmb{\sigma}^*)x^*\| &= \|\mathcal{O}_{\Sigma}(\pmb{\sigma}^*)(\bar{x}+x^*-\bar{x})\| \\
				& \ge \|\mathcal{O}_{\Sigma}(\pmb{\sigma}^*)\bar{x}\|- \|\mathcal{O}_{\Sigma}(\pmb{\sigma}^*)\|x^*-\bar{x}\|\\
				& \ge \underline{\zeta}_k(\omega_N)-\max_{\pmb{\sigma}\in \mathcal{M}^k}\|\mathcal{O}_{\Sigma}(\pmb{\sigma})\| \sqrt{2-2\delta(\frac{\varepsilon M^k}{2})}.
			\end{align*}
			Based on this inequality, we conclude that, with probability no smaller than  $1-(1-\varepsilon)^N$, 
			\begin{align}\label{eqn:minsigmazetak}
				\min_{\pmb{\sigma}\in \mathcal{M}^k}\sigma_{\min}(\mathcal{O}_{\Sigma}(\pmb{\sigma})) \ge& \underline{\zeta}_k(\omega_N) \\
				&- \max_{\pmb{\sigma}\in \mathcal{M}^k}\|\mathcal{O}_{\Sigma}(\pmb{\sigma})\| \sqrt{2-2\delta(\frac{\varepsilon M^k}{2})}. \nonumber
			\end{align}
			\\
			Step 3: We now combine the results from Steps 1 \& 2.  From Step 1, for any $\varepsilon\in (0,1)$, the probability that \eqref{eqn:maxsigmazetak} does not hold is less than $(1-\varepsilon)^N$. From Step 2, for any $\varepsilon'\in (0,1)$ the probability that \eqref{eqn:minsigmazetak} with $\varepsilon'$ does not hold  is also less than $(1-\varepsilon')^N$. Hence, the probability that \eqref{eqn:maxsigmazetak} or \eqref{eqn:minsigmazetak} does not hold becomes $(1-\varepsilon)^N+(1-\varepsilon')^N$, which means that the probability that both of  \eqref{eqn:maxsigmazetak} and \eqref{eqn:minsigmazetak} hold is no smaller than $1-(1-\varepsilon)^N-(1-\varepsilon')^N$. Thus, (\ref{eqn:sigmamaxminpsi}) holds with probability no smaller than $1-(1-\varepsilon)^N-(1-\varepsilon')^N$. $\Box$
			
			From the arguments in the proof of Corollary \ref{cor:muSapriori}, it holds that
			$$
			\chi_{\Sigma}(P,k) \le \sqrt{\left( c_k^2\kappa(P)\right)^{n-1}} 
			$$
			where 
			$$
			c_k = \frac{\max_{\pmb{\sigma}\in \mathcal{M}^k}\sigma_{\max}(\mathcal{O}_{\Sigma}(\pmb{\sigma}))}{\min_{\pmb{\sigma}\in \mathcal{M}^k}\sigma_{\min}(\mathcal{O}_{\Sigma}(\pmb{\sigma})) }.
			$$
			This, together with Theorem \ref{thm:prostability}, and Lemma \ref{lem:normbound}, leads to the result in Theorem \ref{thm!pract}.

			\bibliographystyle{unsrt}
			\bibliography{Reference}
			
		\end{document}